\documentclass[aps,prl,twocolumn,showpacs,showkeys,footinbib,groupedaddress]{revtex4-1}

\usepackage{amsmath}
\usepackage{amssymb}
\usepackage{graphicx}
\usepackage{bm}
\usepackage{color}
\usepackage{relsize}
\usepackage{braket}
\usepackage{bbold}
\usepackage{txfonts}
\usepackage{epstopdf}
\usepackage{hyperref}

\begin{document}
\title{Wireless Majorana Bound States: From Magnetic Tunability to Braiding}
\author{Geoffrey L. Fatin, Alex Matos-Abiague, Benedikt Scharf, and Igor \v{Z}uti\'c}
\affiliation{Department of Physics, University at Buffalo, State University of New York, Buffalo, New York 14260, USA}

\date{\today}

\begin{abstract}
We propose a versatile platform to investigate the existence of Majorana bound states (MBSs) and their non-Abelian statistics through braiding. This implementation combines a two-dimensional electron gas formed in a semiconductor quantum well grown on the surface of an $s$-wave superconductor, with a nearby array of magnetic tunnel junctions (MTJs). The underlying magnetic textures produced by MTJs provide highly-controllable topological phase transitions to confine and transport MBSs in two dimensions, overcoming the requirement for a network of wires. {Obtained scaling relations confirm that various semiconductor quantum well materials are suitable for this proposal}. 
\end{abstract}

\maketitle

In condensed-matter systems Majorana bound states (MBSs) are emergent quasiparticles with exotic non-Abelian statistics and particle-antiparticle symmetry~\cite{Kitaev2001:PU,Wilczek2009:NP,Alicea2012:RPP,Leijnse2012:SST,Beenakker2013:ARCMP,Franz2013:NN,Tkachov2013:PSS}. Elucidating their properties is further motivated by the prospect to use them for fault-tolerant quantum computing~\cite{Nayak2008:RMP,Kitaev2003:AP,Alicea2011:NP}. Typical material systems envisioned for experimental implementations of MBSs include superconducting regions~\footnote{$\nu=5/2$ quantum Hall state [N. Read and D. Green, Phys. Rev. B {\bf 61}, 10267 (2000)] and superfluid $^3\mathrm{He}$ [N.~B. Kopnin and M.~M. Salomaa, Phys. Rev. B {\bf 44}, 9667 (1991)] could provide other implementations.} such as those relying on the native $p$-wave pairing in a vortex core~\cite{Ivanov2001:PRL}, Sr$_2$RuO$_4$~\cite{Mackenzie2003:RMP,ZuticNote}, and Bechgaard salts~\cite{Sengupta2001:PRB}. They can also occur with common proximity-induced $s$-wave pairing when combined with a nontrivial spin structure, which can be provided by spin-orbit coupling (SOC) and magnetic textures~\cite{Fu2008:PRL,Lutchyn2010:PRL,Sau2010:PRL,Oreg2010:PRL,Hasan2010:RMP,Duckheim2011:PRB,Klinovaja2013:PRL,NadjPerge2014:S,Pawlak2015:arxiv} to yield an effective $p$-wave pairing.

Impressive advances in fabricating complex superconducting systems for an unambiguous detection of MBSs remain actively debated~\cite{NadjPerge2014:S,Mourik2012:S,Deng2012:NL,Rokhinson2012:NP,Das2012:NP,Finck2013:PRL,Lee2014:NN}.~Observed MBS signatures, such as a zero-bias tunneling conductance peak, may have other origins~\cite{Liu2012:PRL,Bagrets2012:PRL,Kells2012:PRB,Roy2013:PRB,Peng2015:PRL} and should be supplemented by additional measurements~\cite{DasSarma2012:PRB,Appelbaum2013:APL,Vernek2014:PRB,BenShach2015:PRB,Scharf2015:PRB}. However, those signatures do not directly probe non-Abelian statistics~\cite{Wilczek2009:NP,Alicea2012:RPP,Leijnse2012:SST,Nayak2008:RMP}.~While realizing the non-Abelian braiding statistics under exchange would provide both an ultimate proof for the MBS existence and the key element for topological quantum computing, even theoretical schemes imply a significant complexity to implement such braiding~\cite{Alicea2011:NP}. Exchanging vortices on the surface of the $p$-wave superconductor to close the braiding loop would require an experimental {\em tour de force}. Frequently examined one-dimensional (1D) superconductor/semiconductor wire systems avoid the need for challenging vortex manipulation, but that geometry alone is insufficient. Braiding statistics are ill-defined in 1D and complex wire networks must be used instead of single wires, posing additional obstacles~\cite{Alicea2011:NP,Kim2015:PRB,*Sau2011:PRB,*Clarke2011:PRB,*Halperin2012:PRB,*Klinovaja2013:PRX}.

\begin{figure}[t]
\centering
\includegraphics*[width=8cm]{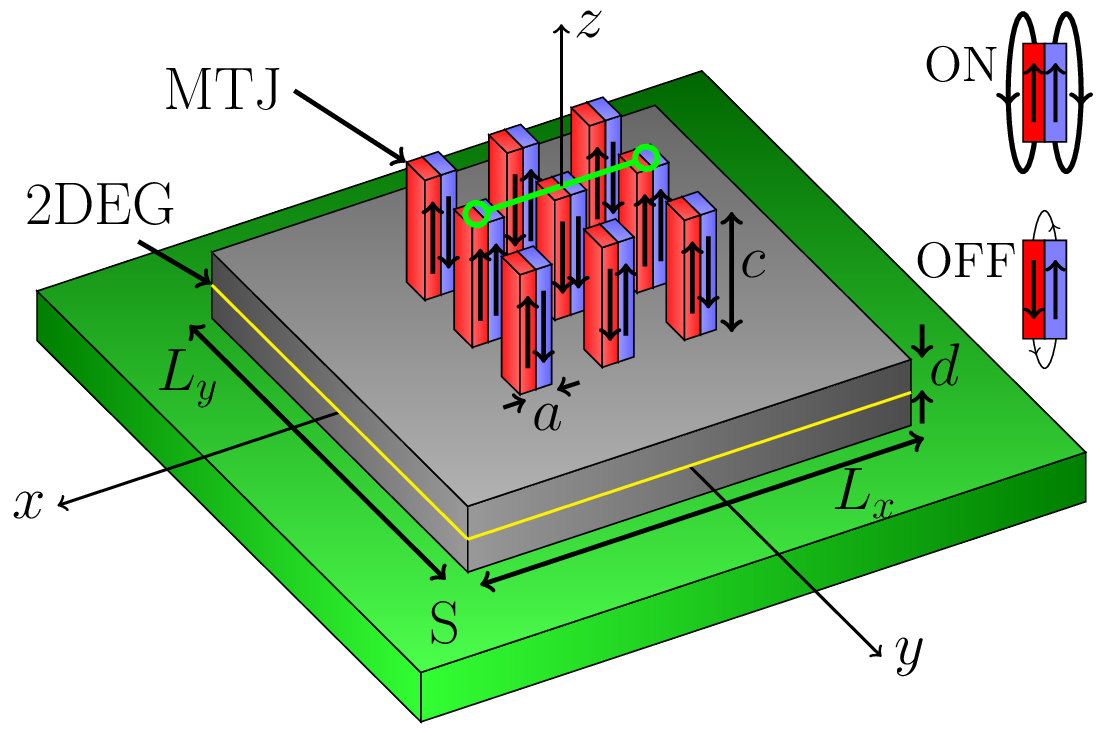}
\caption{Schematic of the setup. A two-dimensional electron gas (2DEG) is formed in a semiconductor quantum well grown on the surface of an $s$-wave superconductor (S). An array of magnetic tunnel junctions (MTJs) produces a magnetic texture, tunable by switching individual MTJs to the parallel (ON) or antiparallel (OFF) configuration. For the depicted array configuration, two Majorana bound states form at the ends of the middle row (green curve).
}\label{fig:Scheme}
\end{figure}

To address these challenges, in Fig.~\ref{fig:Scheme} we propose a versatile platform to realize MBSs and enable their braiding in 2D superconductor/semiconductor systems without the need for wire networks. This proposal seemingly contradicts prior knowledge. In semiconductor wires with SOC-based effective $p$-wave pairing, the energetically isolated MBSs do not survive the transition to 2D, but rather evolve into edge states with increasing wire width~\cite{Potter2010:PRL,Sedlmayr2015:PRB}. Without SOC, a helical magnetic texture can still produce MBSs in 1D systems~\cite{Kjaergaard2012:PRB,Pientka2013:PRB,NadjPerge2014:S}. The MBSs may then survive the transition to 2D, but spread entirely along opposite edges, precluding braiding.

Here we show that in spite of the above observations a properly designed magnetic texture acting on a 2D electron gas (2DEG) with proximity induced $s$-wave superconductivity can support localized MBSs. The effect of the magnetic texture is twofold: (i) it drives local transitions to the topological regime and the emergence of MBSs and (ii) it confines MBSs by forming effective wires. The size, position, and shape of the effective wires can be conveniently modified by altering the magnetic texture, thereby permitting exchange of the MBSs. Remarkably, the required magnetic textures can be generated by an array of magnetic tunnel junctions (MTJs), similar to those used in commercial MTJ-based magnetic random access memory (MRAM)~\cite{Tsymbal2011}. The braiding operation can be realized by a sequence of MTJ switchings in the same way bits are written in MRAMs. Our proposal differs qualitatively from previous schemes, where the lack of tunability of the magnetic texture together with the rigid wire geometry~\cite{Kjaergaard2012:PRB} prevents braiding.

We consider a 2DEG with a large effective $g$-factor, $g^\ast$, lying close to the surface of an $s$-wave superconductor (Fig.~\ref{fig:Scheme}). An array of MTJs creates a magnetic texture in the 2DEG. The system is modeled by using the Bogoliubov-de Gennes (BdG) Hamiltonian,
\begin{equation}\label{hamiltonian}
H = \left(p^2/2m^\ast -\mu \right)\tau_z + \Delta \tau_x + g^\ast \mu_{\rm B} \mathbf{B}
\cdot\boldsymbol{\sigma}/2\;,
\end{equation}
where $\tau_i$ ($\sigma_i$) are the Nambu (Pauli) matrices in particle-hole (spin) space, $\mu_{\rm B}$ is the Bohr magneton, and $\mathbf{p}$ and $m^\ast$ are, respectively, the momentum and effective mass of the carriers. The chemical potential, $\mu$, and the proximity induced superconducting gap, $\Delta$, are assumed to be constant, while $\mathbf{B}$ denotes the inhomogeneous magnetic field generated by the MTJ array. For systems with large $g^\ast$ and moderate magnetic fields the Zeeman term dominates over the orbital effects, neglected in Eq.~(\ref{hamiltonian}).

As schematically shown in Fig.~\ref{fig:Scheme}, each MTJ possesses a hard (blue) layer with a fixed magnetization orientation and a soft (red) layer whose magnetization direction can be flipped in the $yz$-plane. The MTJ is in the ON (OFF) state when the magnetization of the hard and soft layers are parallel (antiparallel) (see Fig.~\ref{fig:Scheme}, right). The specific form of $\mathbf{B}$ depends on the particular state of each, individually switchable, MTJ in the array.

The complexity of the magnetic texture only permits a numerical determination of MBS existence, depending on the specific MTJ array configuration. We therefore solve the eigenvalue problem corresponding to the BdG Hamiltonian in Eq.~(\ref{hamiltonian}) using a fourth order finite-difference scheme~\cite{SM}\nocite{Ramachandran2011:CSE,Chartrand2004,Zudov2002:PRB,Fabian2007:APS}. Nevertheless, one can get some physical insight by locally rotating the spin axes to diagonalize the Zeeman interaction from Eq.~(\ref{hamiltonian})~\cite{Braunecker2010:PRB},
\begin{equation}
\tilde{H} = \left[\left(\mathbf{p}-e\mathbf{A}\right)^2/2m^*-\mu\right]\tau_z + \Delta\tau_x
+ g^\ast \mu_B B\sigma_z/2\;, \: \:
\end{equation}
with the non-Abelian field, $\mathbf{A} = \boldsymbol{\mathcal{A}}(\mathbf{r})\cdot\boldsymbol{\sigma}$, and $\boldsymbol{\mathcal{A}}(\mathbf{r}) =\hbar \left(-\boldsymbol{\nabla}\phi\sin\theta,\boldsymbol{\nabla}\theta,\boldsymbol{\nabla}\phi \cos\theta\right)^T/(2e)$, where $\theta,\phi$ are the spherical coordinates of $\mathbf{B}$. The rotated Hamiltonian $\tilde{H}$ contains the main ingredients for the emergence of MBSs: a proximity-induced superconducting gap, a Zeeman interaction, and an effective SOC resulting from the presence of the non-Abelian field $\mathbf{A}$~\cite{Alicea2010:PRB,Sau2010:PRB}.

\begin{figure}[t]
	\centering
	\includegraphics*[width=8.5cm]{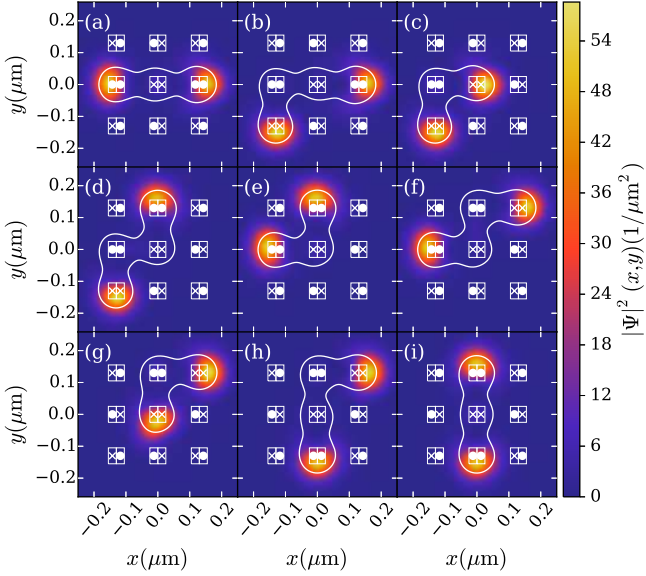}
	\caption{Probability densities of the MBSs corresponding to the initial state (a) and to the first eight stable configurations of the MTJ array [(b) through (i)], $S_0$,...,$S_8$, superimposed with magnetization [white rectangles with dots (crosses) for direction parallel (antiparallel) to the $\mathbf{z}$ axis]. The effective wires are marked with white contours~\cite{SM}.}\label{fig:Probability}
\end{figure}

The formation of topologically nontrivial regions is approximately determined by the following condition,
\begin{equation}\label{t-condition}
\left|\frac{1}{2}\mu_B g^\ast\mathbf{B}\right|^2=\left(\mu - \frac{\hbar^2 \sum_{i=1}^2\frac{\partial \mathbf{B}}{\partial x_i}\cdot \frac{\partial \mathbf{B}}{\partial x_i}}{8 m^\ast|\mathbf{B}|^2}\right)^2+ \Delta^2,
\end{equation}
which locally characterizes the quantum phase transition and can be obtained by following a similar procedure to that of Ref.~\onlinecite{Oreg2010:PRL} (see also Ref.~\onlinecite{SM}). The set of positions {where} Eq.~(\ref{t-condition}) is fulfilled forms a contour that can be used as a guide {to separate} topologically trivial and nontrivial domains. As shown in Fig.~\ref{fig:Probability}, these contours define effective topological wires with MBSs localized at {the} ends.

The proposed MTJ array approach for MBS braiding offers great flexibility in designing the shape and size of the device. For simplicity, we focus on a square $(3\times 3)$ MTJ array, although different arrays can be treated in a similar way~\cite{SM}. This MTJ array may not yield an optimal number of operations needed for exchanging MBSs. However, its compact geometry provides a high scalability and integration with existing MRAM architectures, as well as a flexibility in the manipulation of MBSs.

The hard layer regions are fixed into a checkerboard pattern across the MTJ array, alternating direction between each adjacent site. The set of stable states of a MTJ can be represented as $\{|00\rangle,|01\rangle,|10\rangle,|11\rangle\}$. Here the first and second components of the two-bit states refer to the hard and soft layers, respectively. The bit $|1\rangle$ ($|0\rangle$) refers to a magnetization parallel (antiparallel) to $\hat{\mathbf{z}}$. According to their locations, the MTJs are labeled by $M_{ij}$. Similarly, the magnetic configuration of the MTJ array can be described by a matrix whose elements are the states of the individual MTJs. The system is initialized by bringing the MTJ array into the configuration
\begin{equation}
S_0 = \left(\begin{array}{ccc}
|01\rangle & |10\rangle & |01\rangle \\
|11\rangle & |00\rangle & |11\rangle \\
|01\rangle & |10\rangle & |01\rangle 
\end{array}\right),
\end{equation}
which supports the formation of two MBSs localized beneath $M_{21}$ and $M_{23}$. The braiding operation is then performed by a sequence of MTJ adiabatic switchings which we refer to as the braiding protocol. {The MTJ array configuration $S_n$ at the $n$th step of the protocol can be obtained recursively as $S_n = U_n \circ S_{n-1}$, where $\circ$ denotes a Hadamard product~\cite{Horn1991} and $U_n$ is a $(3\times 3)$ matrix with a NOT gate $u_\pi$ in the position of the MTJ to be switched and unit operators elsewhere. $u_\pi$ acts on soft-layer bits as $u_\pi |m0\rangle = |m1\rangle$, $u_\pi | m1\rangle = |m0\rangle$ for $m\in\{0,1\}$. A detailed explanation of the braiding protocol is given in~\cite{SM}.}

\begin{figure}[t]
	\centering
\includegraphics*[width=8cm]{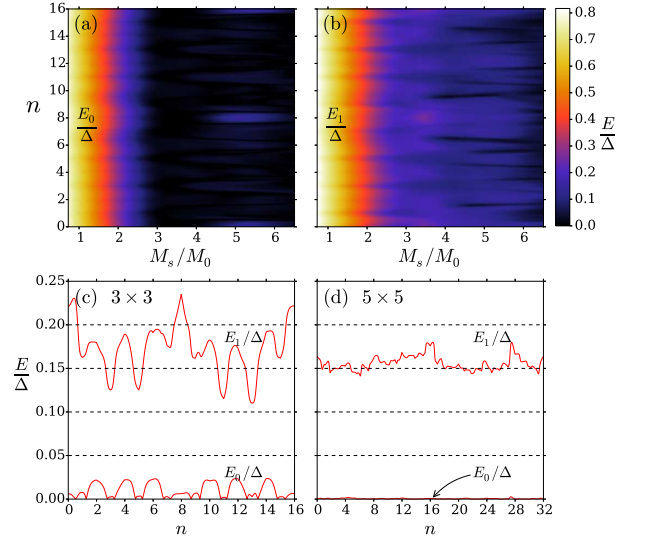}
\caption{{(a) Ground and (b) $1^{\text{st}}$ excited state energies ($E_0$, $E_1$) as functions of saturation magnetization $M_s$ and the $(3\times 3)$ MTJ array configuration $n$. $M_0 = 8\pi \Delta/(g^\ast \mu_B \mu_0)$ with vacuum permeability $\mu_0$. $E_0(n)$, $E_1(n)$ {with $M_s/M_0\approx 3$} for (c) $(3\times 3)$ and (d) $(5\times 5)$ MTJ array.}}
\label{fig:GapsVsMs}
\end{figure}

To illustrate the effect of the braiding protocol on the MBSs, we performed numerical calculations for a 2DEG formed in a (Cd,Mn)Te quantum well~\footnote{Superconducting proximity effects and MBSs in (Cd,Mn) Te are being explored by L. Rokhinson (private comm.).}. Because of its large $g^\ast$ (up to 350) at low temperatures, (Cd,Mn)Te quantum wells have recently been employed in experiments measuring the effects of magnetic textures on the transport properties of novel spin transistors~\cite{Betthausen2012:Science}. Here we use $g^\ast=300$, $m^\ast=0.1~m_0$ ({$m_0$ is the bare electron mass}), $\mu= 0.2~{\rm meV}$, and $\Delta = 0.3~{\rm meV}$. Each MTJ has a square cross section of side length $a=50~{\rm nm}$, and a height of $c=4~a$ (Fig.~\ref{fig:Scheme}). The centers of each MTJ are separated by $2.6~a$ in both the $x$ and $y$ directions. The side lengths of the 2DEG are given by $L_x=L_y=14~a$ and the distance from the nearest edge of each MTJ to the 2DEG is taken to be $d=1.1~a$. The magnetic field $\mathbf{B}$, whose strength is proportional to the saturation magnetization $M_s$ of the MTJ layers, was determined using bound current expressions for each magnetization region~\cite{SM} and assuming $M_s = 1.1\times10^6~{\rm A/m}$. As shown in Fig.~\ref{fig:GapsVsMs}(a), such a value of $M_s$ is nearly optimal for the realization of the braiding protocol for this parameter set.

The probability density of the MBSs for the initial ($S_0$) and first eight configurations ($S_1,...,S_8$) of the MTJ array are shown in Fig.~\ref{fig:Probability}. The white contours derived from Eq.~(\ref{t-condition}) describe the formation of effective, topologically nontrivial quasi-1D wires in the 2D system. The MBSs are strongly localized at the ends of the effective wires, while the corresponding charge distribution is vanishingly small over the entire structure~\cite{SM}. The existence of such effective topological wires strongly depends on the configuration of the MTJ array. By properly changing the magnetic texture, the effective wires can be conveniently modified, resulting in the transport of the MBSs (Fig.~\ref{fig:Probability}) and their exchange. A subsequent set of similar operations is still needed for completing the braiding~\cite{SM}.

While 
Fig.~\ref{fig:Probability} demonstrates the existence of the MBSs at each completed step of the braiding protocol, the MBSs could still fuse during the switching process and prevent the braiding. The switching process of a given MTJ is modeled as an adiabatic rotation in which the relative angle in the $yz$-plane between the magnetizations in the hard and soft layer changes from 0 to $\pi$ (from $\pi$ to 0) during the OFF (ON) operation. The transition to the topological state is determined by the strength of the magnetic field texture $\propto M_s$ of the MTJ layers~\cite{SM}. 

{We therefore investigate the dependence of the ground and first excited state energies ($E_0$, $E_1$) as functions of both $M_s$ [in units of $M_0 = 8\pi\Delta/(g^\ast \mu_{\rm B}\mu_0)$ with vacuum permeability $\mu_0$], and the states of the MTJ array, $n$. The results are shown in Figs.~\ref{fig:GapsVsMs}~(a) and~(b). Integer values of $n$ correspond to stable MTJ configurations, when all of the MTJs are completely switched ON or OFF. Noninteger values of $n$ represent the intermediate states of the MTJ array during the switching process, in which the hard and soft layer magnetizations are not collinear. For example, $0\leq n \leq 1$ corresponds to the adiabatic rotation of the soft layer magnetization of $M_{31}$ during the transition of the MTJ array from state $S_0$ to $S_1$. A topological phase transition is clearly observed at $M_s/M_0\approx 3$ or $M_s \approx 1.1 \times 10^{6} {\rm A}/{\rm m}$. There the MBSs, separated by a finite energy gap $E_1(n)-E_0(n)$ from the first excited state, start to appear [Figs.~\ref{fig:GapsVsMs}~(a) and~(b)]. For $M_s/M_0 \gtrsim 4$, the energy gap closes at some steps of the braiding protocol [Fig.~\ref{fig:GapsVsMs}(a) and (b)], indicating that MBSs fuse into finite energy states. The operational regime for the system used here is therefore limited to $3\lesssim M_s/M_0\lesssim 4$.}

\begin{figure}[t]
	\centering
	\includegraphics[width=8cm]{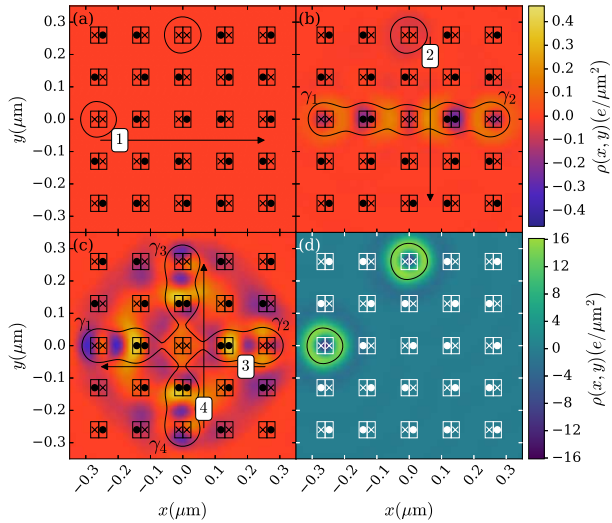}
	\caption{\label{Fig:measurement} {Charge density illustrating the scheme for measuring non-Abelian statistics. (a) The system is initially prepared in a state consisting of two empty QDs. (b) After sequential switching of the MTJs along arrow 1 [see (a)], the left QD evolves into a wire occupied by two end MBSs. (c) After operation 2, the top QD evolves into an occupied wire with another two MBSs. Operations 3 and 4 complete the double braiding and fuse the MBSs. The result is an extra charge in each QD [see (d)].}}
\end{figure}

In Fig.~\ref{fig:GapsVsMs}~(c), $E_0(n)$ and $E_1(n)$ are shown for a $(3\times 3)$ MTJ array at $M_s/M_0\approx 3$. In nonideal systems, overlap between MBSs at opposite ends of a wire will result in a small but finite $E_0$.~Typically, the energy of the MBSs fluctuates {about} zero as a function of system parameters such as size, chemical potential, and magnetic field strength. {Longer effective wires with greater spatial MBS separation imply lower $E_0(n)$} [$E_0(1) \approx E_0(3) \approx E_0(5) \approx E_0(7) < E_0(2) \approx E_0(4) \approx E_0(6)$]~\cite{DasSarma2012:PRB,Lim2012:PRB,*Prada2012:PRB,*Rainis2013:PRB}.
 
The energy isolation of the MBSs from the first excited state is apparent during the full braiding operation. The comparison between Figs.~\ref{fig:GapsVsMs}~(c) and~(d) reveals that increasing the size of the MTJ array to $(5\times 5)$ leads to $E_0(n)$ closer to zero and improves the energy gap by reducing fluctuations of $E_1(n)$. The size of the energy gap in Fig.~\ref{fig:GapsVsMs} corresponds to times $<$ ns, ensuring adiabatic MTJ switching, typically above 10 ns~\cite{Tsymbal2011}. In the worst case scenario (i.e., the smallest energy gap), the switching time should exceed $0.02\rm$ ns for adiabaticity. This condition, readily achieved with MTJs, eliminates the risk of excitation from the ground state during switching. While we have focused on (Cd,Mn)Te, using scaling analysis~\cite{SM} our findings can be applied to other materials systems with large $g$-factors in which the superconducting proximity effects were already measured~\cite{Mourik2012:S,Hart2015:arXiv,*Xu2014:NP,*Shabani2015:PRB}, {or to In(As,Sb) which is promising for MBSs \cite{Winkler2016:arxiv}}. Denoting the new values with tildes, $\lambda_g=\tilde{g}^\ast/g^\ast$, $\lambda_m= \tilde{m}^\ast/m^\ast$, the spatial dimensions rescaled by $\lambda_a=1/\sqrt{|\lambda_g| \lambda_m}$ yield results identical to Fig.~\ref{fig:GapsVsMs}. For example, in $n$-doped (In,Mn)As with $\tilde{g}^\ast=100$~\cite{Zutic2004:RMP} and $\tilde{m}^\ast=0.026~m_0$, $\lambda_g=1/3$, $\lambda_m=13/50$, and $\lambda_a\approx 3.4$. {Our scaling results are also preserved with additional SOC \cite{SM}}.

In addition to allowing for the manipulation of MBSs, the proposed system could also be used for investigating the signatures of those states on transport properties. For example, by attaching top contacts at the ends of a topological wire corresponding to a given MTJ array configuration, the presence of MBSs should yield a zero-bias peak in the differential conductance. Such a peak would be modified and eventually suppressed~\cite{Lin2012:PRB, DasSarma2012:PRB}, once the effective topological wire is modified in such a way that its ends are no longer near the contacts. Alternatively, scanning tunneling microscopy (STM) measurements can also be employed to probe the MBSs at the ends of the effective wires. In our scheme, at least one MBS is always localized below an outer MTJ of the array and can therefore be accessed by a lateral STM technique~\cite{Fridman2011:APL}.

{Beyond manipulations and detection of MBSs, the ultimate goal is to probe the non-Abelian statistics, for which at least 4 MBSs are needed. In Fig.~\ref{Fig:measurement}, we show a possible scheme for probing non-Abelian statistics by local charge measurements. The system is initially prepared as in Fig.~\ref{Fig:measurement}(a), where two empty effective quantum dots (QDs) { are formed on the left (top) of the MTJ array, represented separately by $|0\rangle_\alpha$ with $\alpha = l~(\alpha=t)$ and together by the product state $|00\rangle = |0\rangle_l \otimes |0\rangle_t$}. Note that in such a configuration there are no MBSs. As the MTJ array configuration adiabatically evolves from (a) to (b), the left QD transforms into a topological wire occupied by two end MBSs (the charge remains small because the MBSs are chargeless). Another pair of chargeless MBSs occupy the topological wire resulting from the elongation of the top QD [see Fig.~\ref{Fig:measurement}(c)]. {The fermionic creation (annihilation) operator of each QD is given by $\hat{f}_\alpha^\dagger$ ($\hat{f}_\alpha$) for $\alpha = l,t$. The MBSs are represented by operators $\hat{\gamma}_1 = \hat{f}_l^\dagger + \hat{f}_l$, $\hat{\gamma}_2 = i(\hat{f}^\dagger_l - \hat{f}_l)$, $\hat{\gamma}_3 = \hat{f}_t^\dagger + \hat{f}_t$, and $\hat{\gamma}_4 = i(\hat{f}^\dagger_t - \hat{f}_t)$~\cite{Kitaev2001:PU}. Steps 1, 2 correspond to a counterclockwise exchange of $\gamma_2,\gamma_4$ represented by the braiding operator $\hat{B}_{24} = (1-\hat{\gamma}_2 \hat{\gamma}_4)/\sqrt{2}$ \cite{Leijnse2012:SST}.} The MBSs subsequently fuse after returning the topological wires to the initial effective QDs, according to operations 3 and 4 shown in (c) { which again exchange $\gamma_2,\gamma_4$.} This results in the appearance of extra charge in each QD (d) {(note the change in scale by a factor of nearly 40)} {since
\begin{equation}
\hat{B}_{24}^2 |00\rangle = -\hat{\gamma}_2 \hat{\gamma}_4 |00\rangle = |11\rangle \;.
\end{equation}
The order of operations is crucial, as double braiding is realized in the sequence shown in Fig.~\ref{Fig:measurement} but not if the order of operations 3 and 4 is reversed.} In this case, {$\hat{B}_{42}\hat{B}_{24} |00\rangle = |00\rangle$ and} the final state will correspond to empty rather than charged QDs. Thus, by contrasting the initial and final charges of the effective QDs (this can be measured by using single-electron transistor spectroscopy \cite{Ilani2004:Nature,Martin2004:Science} or lateral STM \cite{Fridman2011:APL}) one can probe the non-Abelian statistics \cite{SM}.}

We have shown that for a two-dimensional electron gas under the influence of an $s$-wave superconductor and different configurations of a magnetic tunnel junction array, Majorana bound states are possible {for various semiconductor quantum well materials}. In future work using this scheme, it would be interesting to explore the interplay between Majorana bound states and transport, as well as their stability. These effects, studied in other systems, have already shown a highly-nontrivial behavior~\cite{Asano2010:PRL,Gangadharaiah2011:PRL,Stoudenmire2011:PRB,Nakosai2013:PRL,Adagideli2014:PRB,Valentini2014:PRB,Kashuba2015:PRL}. The proposed magnetic tunnel junction arrays offer great tunability of attainable magnetic configurations to accurately reposition Majorana states using only effective wires. Our protocol for the exchange of Majorana states provides a method for investigating their non-Abelian statistics and other novel applications~\cite{Kovalev2014:PRL}.

\noindent\emph{Acknowledgments.} We thank A. Kent, L. Rokhinson, N. Samarth, J. Shabani, J.~Y.~T. Wei, T. Wojtowicz, and A. Yacoby for valuable discussions. This work was supported by U.S. ONR Grant N000141310754 (G.F., B.S.), U.S. DOE, Office of Science BES, under Award DE-SC0004890 (A.M.-A., I.\v{Z}), DFG Grant No. SCHA 1899/1-1 (B.S.), and NSF ECCS-1508873 (G.F.).


\bibliographystyle{apsrev4-1}
\bibliography{BibTopInsAndTopSup_v14}

\begin{thebibliography}{81}%
\makeatletter
\providecommand \@ifxundefined [1]{%
 \@ifx{#1\undefined}
}%
\providecommand \@ifnum [1]{%
 \ifnum #1\expandafter \@firstoftwo
 \else \expandafter \@secondoftwo
 \fi
}%
\providecommand \@ifx [1]{%
 \ifx #1\expandafter \@firstoftwo
 \else \expandafter \@secondoftwo
 \fi
}%
\providecommand \natexlab [1]{#1}%
\providecommand \enquote  [1]{``#1''}%
\providecommand \bibnamefont  [1]{#1}%
\providecommand \bibfnamefont [1]{#1}%
\providecommand \citenamefont [1]{#1}%
\providecommand \href@noop [0]{\@secondoftwo}%
\providecommand \href [0]{\begingroup \@sanitize@url \@href}%
\providecommand \@href[1]{\@@startlink{#1}\@@href}%
\providecommand \@@href[1]{\endgroup#1\@@endlink}%
\providecommand \@sanitize@url [0]{\catcode `\\12\catcode `\$12\catcode
  `\&12\catcode `\#12\catcode `\^12\catcode `\_12\catcode `\%12\relax}%
\providecommand \@@startlink[1]{}%
\providecommand \@@endlink[0]{}%
\providecommand \url  [0]{\begingroup\@sanitize@url \@url }%
\providecommand \@url [1]{\endgroup\@href {#1}{\urlprefix }}%
\providecommand \urlprefix  [0]{URL }%
\providecommand \Eprint [0]{\href }%
\providecommand \doibase [0]{http://dx.doi.org/}%
\providecommand \selectlanguage [0]{\@gobble}%
\providecommand \bibinfo  [0]{\@secondoftwo}%
\providecommand \bibfield  [0]{\@secondoftwo}%
\providecommand \translation [1]{[#1]}%
\providecommand \BibitemOpen [0]{}%
\providecommand \bibitemStop [0]{}%
\providecommand \bibitemNoStop [0]{.\EOS\space}%
\providecommand \EOS [0]{\spacefactor3000\relax}%
\providecommand \BibitemShut  [1]{\csname bibitem#1\endcsname}%
\let\auto@bib@innerbib\@empty
\bibitem [{\citenamefont {Kitaev}(2001)}]{Kitaev2001:PU}%
  \BibitemOpen
  \bibfield  {author} {\bibinfo {author} {\bibfnamefont {A.~Y.}\ \bibnamefont
  {Kitaev}},\ }\href@noop {} {\bibfield  {journal} {\bibinfo  {journal}
  {Phys.-Usp.}\ }\textbf {\bibinfo {volume} {44}},\ \bibinfo {pages} {131}
  (\bibinfo {year} {2001})}\BibitemShut {NoStop}%
\bibitem [{\citenamefont {Wilczek}(2009)}]{Wilczek2009:NP}%
  \BibitemOpen
  \bibfield  {author} {\bibinfo {author} {\bibfnamefont {F.}~\bibnamefont
  {Wilczek}},\ }\href {\doibase 10.1038/nphys1380} {\bibfield  {journal}
  {\bibinfo  {journal} {Nat. Phys.}\ }\textbf {\bibinfo {volume} {5}},\
  \bibinfo {pages} {614} (\bibinfo {year} {2009})}\BibitemShut {NoStop}%
\bibitem [{\citenamefont {Alicea}(2012)}]{Alicea2012:RPP}%
  \BibitemOpen
  \bibfield  {author} {\bibinfo {author} {\bibfnamefont {J.}~\bibnamefont
  {Alicea}},\ }\href@noop {} {\bibfield  {journal} {\bibinfo  {journal} {Rep.
  Prog. Phys.}\ }\textbf {\bibinfo {volume} {75}},\ \bibinfo {pages} {076501}
  (\bibinfo {year} {2012})}\BibitemShut {NoStop}%
\bibitem [{\citenamefont {Leijnse}\ and\ \citenamefont
  {Flensberg}(2012)}]{Leijnse2012:SST}%
  \BibitemOpen
  \bibfield  {author} {\bibinfo {author} {\bibfnamefont {M.}~\bibnamefont
  {Leijnse}}\ and\ \bibinfo {author} {\bibfnamefont {K.}~\bibnamefont
  {Flensberg}},\ }\href@noop {} {\bibfield  {journal} {\bibinfo  {journal}
  {Semicond. Sci. and Technol.}\ }\textbf {\bibinfo {volume} {27}},\ \bibinfo
  {pages} {124003} (\bibinfo {year} {2012})}\BibitemShut {NoStop}%
\bibitem [{\citenamefont {Beenakker}(2013)}]{Beenakker2013:ARCMP}%
  \BibitemOpen
  \bibfield  {author} {\bibinfo {author} {\bibfnamefont {C.}~\bibnamefont
  {Beenakker}},\ }\href {\doibase 10.1146/annurev-conmatphys-030212-184337}
  {\bibfield  {journal} {\bibinfo  {journal} {Annu. Rev. Condens. Matter
  Phys.}\ }\textbf {\bibinfo {volume} {4}},\ \bibinfo {pages} {113} (\bibinfo
  {year} {2013})}\BibitemShut {NoStop}%
\bibitem [{\citenamefont {Franz}(2013)}]{Franz2013:NN}%
  \BibitemOpen
  \bibfield  {author} {\bibinfo {author} {\bibfnamefont {M.}~\bibnamefont
  {Franz}},\ }\href {\doibase 10.1038/nnano.2013.33} {\bibfield  {journal}
  {\bibinfo  {journal} {Nat. Nanotechnol.}\ }\textbf {\bibinfo {volume} {8}},\
  \bibinfo {pages} {149} (\bibinfo {year} {2013})}\BibitemShut {NoStop}%
\bibitem [{\citenamefont {Tkachov}\ and\ \citenamefont
  {Hankiewicz}(2013)}]{Tkachov2013:PSS}%
  \BibitemOpen
  \bibfield  {author} {\bibinfo {author} {\bibfnamefont {G.}~\bibnamefont
  {Tkachov}}\ and\ \bibinfo {author} {\bibfnamefont {E.~M.}\ \bibnamefont
  {Hankiewicz}},\ }\href {\doibase 10.1002/pssb.201248385} {\bibfield
  {journal} {\bibinfo  {journal} {Phys. Status Solidi B}\ }\textbf {\bibinfo
  {volume} {250}},\ \bibinfo {pages} {215} (\bibinfo {year}
  {2013})}\BibitemShut {NoStop}%
\bibitem [{\citenamefont {Nayak}\ \emph {et~al.}(2008)\citenamefont {Nayak},
  \citenamefont {Simon}, \citenamefont {Stern}, \citenamefont {Freedman},\ and\
  \citenamefont {Das~Sarma}}]{Nayak2008:RMP}%
  \BibitemOpen
  \bibfield  {author} {\bibinfo {author} {\bibfnamefont {C.}~\bibnamefont
  {Nayak}}, \bibinfo {author} {\bibfnamefont {S.~H.}\ \bibnamefont {Simon}},
  \bibinfo {author} {\bibfnamefont {A.}~\bibnamefont {Stern}}, \bibinfo
  {author} {\bibfnamefont {M.}~\bibnamefont {Freedman}}, \ and\ \bibinfo
  {author} {\bibfnamefont {S.}~\bibnamefont {Das~Sarma}},\ }\href {\doibase
  10.1103/RevModPhys.80.1083} {\bibfield  {journal} {\bibinfo  {journal} {Rev.
  Mod. Phys.}\ }\textbf {\bibinfo {volume} {80}},\ \bibinfo {pages} {1083}
  (\bibinfo {year} {2008})}\BibitemShut {NoStop}%
\bibitem [{\citenamefont {Kitaev}(2003)}]{Kitaev2003:AP}%
  \BibitemOpen
  \bibfield  {author} {\bibinfo {author} {\bibfnamefont {A.}~\bibnamefont
  {Kitaev}},\ }\href {http://stacks.iop.org/1468-6996/15/i=6/a=064402}
  {\bibfield  {journal} {\bibinfo  {journal} {Ann. Phys.}\ }\textbf {\bibinfo
  {volume} {303}},\ \bibinfo {pages} {2} (\bibinfo {year} {2003})}\BibitemShut
  {NoStop}%
\bibitem [{\citenamefont {Alicea}\ \emph {et~al.}(2011)\citenamefont {Alicea},
  \citenamefont {Oreg}, \citenamefont {Refael}, \citenamefont {von Oppen},\
  and\ \citenamefont {Fisher}}]{Alicea2011:NP}%
  \BibitemOpen
  \bibfield  {author} {\bibinfo {author} {\bibfnamefont {J.}~\bibnamefont
  {Alicea}}, \bibinfo {author} {\bibfnamefont {Y.}~\bibnamefont {Oreg}},
  \bibinfo {author} {\bibfnamefont {G.}~\bibnamefont {Refael}}, \bibinfo
  {author} {\bibfnamefont {F.}~\bibnamefont {von Oppen}}, \ and\ \bibinfo
  {author} {\bibfnamefont {M.~P.~A.}\ \bibnamefont {Fisher}},\ }\href {\doibase
  10.1038/nphys1915} {\bibfield  {journal} {\bibinfo  {journal} {Nat. Phys.}\
  }\textbf {\bibinfo {volume} {7}},\ \bibinfo {pages} {412} (\bibinfo {year}
  {2011})}\BibitemShut {NoStop}%
\bibitem [{Note1()}]{Note1}%
  \BibitemOpen
  \bibinfo {note} {$\nu =5/2$ quantum Hall state [N. Read and D. Green, Phys.
  Rev. B {\protect \bf 61}, 10267 (2000)] and superfluid $^3\protect \mathrm
  {He}$ [N.~B. Kopnin and M.~M. Salomaa, Phys. Rev. B {\protect \bf 44}, 9667
  (1991)] could provide other implementations.}\BibitemShut {Stop}%
\bibitem [{\citenamefont {Ivanov}(2001)}]{Ivanov2001:PRL}%
  \BibitemOpen
  \bibfield  {author} {\bibinfo {author} {\bibfnamefont {D.~A.}\ \bibnamefont
  {Ivanov}},\ }\href {\doibase 10.1103/PhysRevLett.86.268} {\bibfield
  {journal} {\bibinfo  {journal} {Phys. Rev. Lett.}\ }\textbf {\bibinfo
  {volume} {86}},\ \bibinfo {pages} {268} (\bibinfo {year} {2001})}\BibitemShut
  {NoStop}%
\bibitem [{\citenamefont {Mackenzie}\ and\ \citenamefont
  {Maeno}(2003)}]{Mackenzie2003:RMP}%
  \BibitemOpen
  \bibfield  {author} {\bibinfo {author} {\bibfnamefont {A.~P.}\ \bibnamefont
  {Mackenzie}}\ and\ \bibinfo {author} {\bibfnamefont {Y.}~\bibnamefont
  {Maeno}},\ }\href {\doibase 10.1103/RevModPhys.75.657} {\bibfield  {journal}
  {\bibinfo  {journal} {Rev. Mod. Phys.}\ }\textbf {\bibinfo {volume} {75}},\
  \bibinfo {pages} {657} (\bibinfo {year} {2003})}\BibitemShut {NoStop}%
\bibitem [{Zut()}]{ZuticNote}%
  \BibitemOpen
  \href@noop {} {}\bibinfo {note} {However, there are also possible alternative
  scenarios for chiral pairing symmetry [I. \v{Z}uti\'c and I. Mazin, Phys.
  Rev. Lett. {\bf 95}, 217004 (2005)]}\BibitemShut {NoStop}%
\bibitem [{\citenamefont {Sengupta}\ \emph {et~al.}(2001)\citenamefont
  {Sengupta}, \citenamefont {\ifmmode \check{Z}\else
  \v{Z}\fi{}uti\ifmmode~\acute{c}\else \'{c}\fi{}}, \citenamefont {Kwon},
  \citenamefont {Yakovenko},\ and\ \citenamefont
  {Das~Sarma}}]{Sengupta2001:PRB}%
  \BibitemOpen
  \bibfield  {author} {\bibinfo {author} {\bibfnamefont {K.}~\bibnamefont
  {Sengupta}}, \bibinfo {author} {\bibfnamefont {I.}~\bibnamefont {\ifmmode
  \check{Z}\else \v{Z}\fi{}uti\ifmmode~\acute{c}\else \'{c}\fi{}}}, \bibinfo
  {author} {\bibfnamefont {H.-J.}\ \bibnamefont {Kwon}}, \bibinfo {author}
  {\bibfnamefont {V.~M.}\ \bibnamefont {Yakovenko}}, \ and\ \bibinfo {author}
  {\bibfnamefont {S.}~\bibnamefont {Das~Sarma}},\ }\href {\doibase
  10.1103/PhysRevB.63.144531} {\bibfield  {journal} {\bibinfo  {journal} {Phys.
  Rev. B}\ }\textbf {\bibinfo {volume} {63}},\ \bibinfo {pages} {144531}
  (\bibinfo {year} {2001})}\BibitemShut {NoStop}%
\bibitem [{\citenamefont {Fu}\ and\ \citenamefont {Kane}(2008)}]{Fu2008:PRL}%
  \BibitemOpen
  \bibfield  {author} {\bibinfo {author} {\bibfnamefont {L.}~\bibnamefont
  {Fu}}\ and\ \bibinfo {author} {\bibfnamefont {C.~L.}\ \bibnamefont {Kane}},\
  }\href {\doibase 10.1103/PhysRevLett.100.096407} {\bibfield  {journal}
  {\bibinfo  {journal} {Phys. Rev. Lett.}\ }\textbf {\bibinfo {volume} {100}},\
  \bibinfo {pages} {096407} (\bibinfo {year} {2008})}\BibitemShut {NoStop}%
\bibitem [{\citenamefont {Lutchyn}\ \emph {et~al.}(2010)\citenamefont
  {Lutchyn}, \citenamefont {Sau},\ and\ \citenamefont
  {Das~Sarma}}]{Lutchyn2010:PRL}%
  \BibitemOpen
  \bibfield  {author} {\bibinfo {author} {\bibfnamefont {R.~M.}\ \bibnamefont
  {Lutchyn}}, \bibinfo {author} {\bibfnamefont {J.~D.}\ \bibnamefont {Sau}}, \
  and\ \bibinfo {author} {\bibfnamefont {S.}~\bibnamefont {Das~Sarma}},\ }\href
  {\doibase 10.1103/PhysRevLett.105.077001} {\bibfield  {journal} {\bibinfo
  {journal} {Phys. Rev. Lett.}\ }\textbf {\bibinfo {volume} {105}},\ \bibinfo
  {pages} {077001} (\bibinfo {year} {2010})}\BibitemShut {NoStop}%
\bibitem [{\citenamefont {Sau}\ \emph {et~al.}(2010{\natexlab{a}})\citenamefont
  {Sau}, \citenamefont {Lutchyn}, \citenamefont {Tewari},\ and\ \citenamefont
  {Das~Sarma}}]{Sau2010:PRL}%
  \BibitemOpen
  \bibfield  {author} {\bibinfo {author} {\bibfnamefont {J.~D.}\ \bibnamefont
  {Sau}}, \bibinfo {author} {\bibfnamefont {R.~M.}\ \bibnamefont {Lutchyn}},
  \bibinfo {author} {\bibfnamefont {S.}~\bibnamefont {Tewari}}, \ and\ \bibinfo
  {author} {\bibfnamefont {S.}~\bibnamefont {Das~Sarma}},\ }\href {\doibase
  10.1103/PhysRevLett.104.040502} {\bibfield  {journal} {\bibinfo  {journal}
  {Phys. Rev. Lett.}\ }\textbf {\bibinfo {volume} {104}},\ \bibinfo {pages}
  {040502} (\bibinfo {year} {2010}{\natexlab{a}})}\BibitemShut {NoStop}%
\bibitem [{\citenamefont {Oreg}\ \emph {et~al.}(2010)\citenamefont {Oreg},
  \citenamefont {Refael},\ and\ \citenamefont {von Oppen}}]{Oreg2010:PRL}%
  \BibitemOpen
  \bibfield  {author} {\bibinfo {author} {\bibfnamefont {Y.}~\bibnamefont
  {Oreg}}, \bibinfo {author} {\bibfnamefont {G.}~\bibnamefont {Refael}}, \ and\
  \bibinfo {author} {\bibfnamefont {F.}~\bibnamefont {von Oppen}},\ }\href
  {\doibase 10.1103/PhysRevLett.105.177002} {\bibfield  {journal} {\bibinfo
  {journal} {Phys. Rev. Lett.}\ }\textbf {\bibinfo {volume} {105}},\ \bibinfo
  {pages} {177002} (\bibinfo {year} {2010})}\BibitemShut {NoStop}%
\bibitem [{\citenamefont {Hasan}\ and\ \citenamefont
  {Kane}(2010)}]{Hasan2010:RMP}%
  \BibitemOpen
  \bibfield  {author} {\bibinfo {author} {\bibfnamefont {M.~Z.}\ \bibnamefont
  {Hasan}}\ and\ \bibinfo {author} {\bibfnamefont {C.~L.}\ \bibnamefont
  {Kane}},\ }\href {\doibase 10.1103/RevModPhys.82.3045} {\bibfield  {journal}
  {\bibinfo  {journal} {Rev. Mod. Phys.}\ }\textbf {\bibinfo {volume} {82}},\
  \bibinfo {pages} {3045} (\bibinfo {year} {2010})}\BibitemShut {NoStop}%
\bibitem [{\citenamefont {Duckheim}\ and\ \citenamefont
  {Brouwer}(2011)}]{Duckheim2011:PRB}%
  \BibitemOpen
  \bibfield  {author} {\bibinfo {author} {\bibfnamefont {M.}~\bibnamefont
  {Duckheim}}\ and\ \bibinfo {author} {\bibfnamefont {P.~W.}\ \bibnamefont
  {Brouwer}},\ }\href {\doibase 10.1103/PhysRevB.83.054513} {\bibfield
  {journal} {\bibinfo  {journal} {Phys. Rev. B}\ }\textbf {\bibinfo {volume}
  {83}},\ \bibinfo {pages} {054513} (\bibinfo {year} {2011})}\BibitemShut
  {NoStop}%
\bibitem [{\citenamefont {Klinovaja}\ \emph {et~al.}(2013)\citenamefont
  {Klinovaja}, \citenamefont {Stano}, \citenamefont {Yazdani},\ and\
  \citenamefont {Loss}}]{Klinovaja2013:PRL}%
  \BibitemOpen
  \bibfield  {author} {\bibinfo {author} {\bibfnamefont {J.}~\bibnamefont
  {Klinovaja}}, \bibinfo {author} {\bibfnamefont {P.}~\bibnamefont {Stano}},
  \bibinfo {author} {\bibfnamefont {A.}~\bibnamefont {Yazdani}}, \ and\
  \bibinfo {author} {\bibfnamefont {D.}~\bibnamefont {Loss}},\ }\href {\doibase
  10.1103/PhysRevLett.111.186805} {\bibfield  {journal} {\bibinfo  {journal}
  {Phys. Rev. Lett.}\ }\textbf {\bibinfo {volume} {111}},\ \bibinfo {pages}
  {186805} (\bibinfo {year} {2013})}\BibitemShut {NoStop}%
\bibitem [{\citenamefont {Nadj-Perge}\ \emph {et~al.}(2014)\citenamefont
  {Nadj-Perge}, \citenamefont {Drozdov}, \citenamefont {Li}, \citenamefont
  {Chen}, \citenamefont {Jeon}, \citenamefont {Seo}, \citenamefont {MacDonald},
  \citenamefont {Bernevig},\ and\ \citenamefont {Yazdani}}]{NadjPerge2014:S}%
  \BibitemOpen
  \bibfield  {author} {\bibinfo {author} {\bibfnamefont {S.}~\bibnamefont
  {Nadj-Perge}}, \bibinfo {author} {\bibfnamefont {I.~K.}\ \bibnamefont
  {Drozdov}}, \bibinfo {author} {\bibfnamefont {J.}~\bibnamefont {Li}},
  \bibinfo {author} {\bibfnamefont {H.}~\bibnamefont {Chen}}, \bibinfo {author}
  {\bibfnamefont {S.}~\bibnamefont {Jeon}}, \bibinfo {author} {\bibfnamefont
  {J.}~\bibnamefont {Seo}}, \bibinfo {author} {\bibfnamefont {A.~H.}\
  \bibnamefont {MacDonald}}, \bibinfo {author} {\bibfnamefont {B.~A.}\
  \bibnamefont {Bernevig}}, \ and\ \bibinfo {author} {\bibfnamefont
  {A.}~\bibnamefont {Yazdani}},\ }\href {\doibase 10.1126/science.1259327}
  {\bibfield  {journal} {\bibinfo  {journal} {Science}\ }\textbf {\bibinfo
  {volume} {346}},\ \bibinfo {pages} {602} (\bibinfo {year}
  {2014})}\BibitemShut {NoStop}%
\bibitem [{\citenamefont {{Pawlak}}\ \emph {et~al.}(2015)\citenamefont
  {{Pawlak}}, \citenamefont {{Kisiel}}, \citenamefont {{Klinovaja}},
  \citenamefont {{Meier}}, \citenamefont {{Kawai}}, \citenamefont {{Glatzel}},
  \citenamefont {{Loss}},\ and\ \citenamefont {{Meyer}}}]{Pawlak2015:arxiv}%
  \BibitemOpen
  \bibfield  {author} {\bibinfo {author} {\bibfnamefont {R.}~\bibnamefont
  {{Pawlak}}}, \bibinfo {author} {\bibfnamefont {M.}~\bibnamefont {{Kisiel}}},
  \bibinfo {author} {\bibfnamefont {J.}~\bibnamefont {{Klinovaja}}}, \bibinfo
  {author} {\bibfnamefont {T.}~\bibnamefont {{Meier}}}, \bibinfo {author}
  {\bibfnamefont {S.}~\bibnamefont {{Kawai}}}, \bibinfo {author} {\bibfnamefont
  {T.}~\bibnamefont {{Glatzel}}}, \bibinfo {author} {\bibfnamefont
  {D.}~\bibnamefont {{Loss}}}, \ and\ \bibinfo {author} {\bibfnamefont
  {E.}~\bibnamefont {{Meyer}}},\ }\href@noop {} {\bibfield  {journal} {\bibinfo
   {journal} {arXiv:1505.06078}\ } (\bibinfo {year} {2015})}\BibitemShut
  {NoStop}%
\bibitem [{\citenamefont {Mourik}\ \emph {et~al.}(2012)\citenamefont {Mourik},
  \citenamefont {Zuo}, \citenamefont {Frolov}, \citenamefont {Plissard},
  \citenamefont {Bakkers},\ and\ \citenamefont {Kouwenhoven}}]{Mourik2012:S}%
  \BibitemOpen
  \bibfield  {author} {\bibinfo {author} {\bibfnamefont {V.}~\bibnamefont
  {Mourik}}, \bibinfo {author} {\bibfnamefont {K.}~\bibnamefont {Zuo}},
  \bibinfo {author} {\bibfnamefont {S.~M.}\ \bibnamefont {Frolov}}, \bibinfo
  {author} {\bibfnamefont {S.~R.}\ \bibnamefont {Plissard}}, \bibinfo {author}
  {\bibfnamefont {E.~P. A.~M.}\ \bibnamefont {Bakkers}}, \ and\ \bibinfo
  {author} {\bibfnamefont {L.~P.}\ \bibnamefont {Kouwenhoven}},\ }\href@noop {}
  {\bibfield  {journal} {\bibinfo  {journal} {Science}\ }\textbf {\bibinfo
  {volume} {336}},\ \bibinfo {pages} {1003} (\bibinfo {year}
  {2012})}\BibitemShut {NoStop}%
\bibitem [{\citenamefont {Deng}\ \emph {et~al.}(2012)\citenamefont {Deng},
  \citenamefont {Yu}, \citenamefont {Huang}, \citenamefont {Larsson},
  \citenamefont {Caroff},\ and\ \citenamefont {Xu}}]{Deng2012:NL}%
  \BibitemOpen
  \bibfield  {author} {\bibinfo {author} {\bibfnamefont {M.~T.}\ \bibnamefont
  {Deng}}, \bibinfo {author} {\bibfnamefont {C.~L.}\ \bibnamefont {Yu}},
  \bibinfo {author} {\bibfnamefont {G.~Y.}\ \bibnamefont {Huang}}, \bibinfo
  {author} {\bibfnamefont {M.}~\bibnamefont {Larsson}}, \bibinfo {author}
  {\bibfnamefont {P.}~\bibnamefont {Caroff}}, \ and\ \bibinfo {author}
  {\bibfnamefont {H.~Q.}\ \bibnamefont {Xu}},\ }\href {\doibase
  10.1021/nl303758w} {\bibfield  {journal} {\bibinfo  {journal} {Nano Letters}\
  }\textbf {\bibinfo {volume} {12}},\ \bibinfo {pages} {6414} (\bibinfo {year}
  {2012})}\BibitemShut {NoStop}%
\bibitem [{\citenamefont {Rokhinson}\ \emph {et~al.}(2012)\citenamefont
  {Rokhinson}, \citenamefont {Liu},\ and\ \citenamefont
  {Furdyna}}]{Rokhinson2012:NP}%
  \BibitemOpen
  \bibfield  {author} {\bibinfo {author} {\bibfnamefont {L.~P.}\ \bibnamefont
  {Rokhinson}}, \bibinfo {author} {\bibfnamefont {X.}~\bibnamefont {Liu}}, \
  and\ \bibinfo {author} {\bibfnamefont {J.~K.}\ \bibnamefont {Furdyna}},\
  }\href {\doibase 10.1038/nphys2429} {\bibfield  {journal} {\bibinfo
  {journal} {Nat. Phys.}\ }\textbf {\bibinfo {volume} {8}},\ \bibinfo {pages}
  {795} (\bibinfo {year} {2012})}\BibitemShut {NoStop}%
\bibitem [{\citenamefont {Das}\ \emph {et~al.}(2012)\citenamefont {Das},
  \citenamefont {Ronen}, \citenamefont {Most}, \citenamefont {Oreg},
  \citenamefont {Heiblum},\ and\ \citenamefont {Shtrikman}}]{Das2012:NP}%
  \BibitemOpen
  \bibfield  {author} {\bibinfo {author} {\bibfnamefont {A.}~\bibnamefont
  {Das}}, \bibinfo {author} {\bibfnamefont {Y.}~\bibnamefont {Ronen}}, \bibinfo
  {author} {\bibfnamefont {Y.}~\bibnamefont {Most}}, \bibinfo {author}
  {\bibfnamefont {Y.}~\bibnamefont {Oreg}}, \bibinfo {author} {\bibfnamefont
  {M.}~\bibnamefont {Heiblum}}, \ and\ \bibinfo {author} {\bibfnamefont
  {H.}~\bibnamefont {Shtrikman}},\ }\href {\doibase 10.1038/nphys2479}
  {\bibfield  {journal} {\bibinfo  {journal} {Nat. Phys.}\ }\textbf {\bibinfo
  {volume} {8}},\ \bibinfo {pages} {887} (\bibinfo {year} {2012})}\BibitemShut
  {NoStop}%
\bibitem [{\citenamefont {Finck}\ \emph {et~al.}(2013)\citenamefont {Finck},
  \citenamefont {Van~Harlingen}, \citenamefont {Mohseni}, \citenamefont
  {Jung},\ and\ \citenamefont {Li}}]{Finck2013:PRL}%
  \BibitemOpen
  \bibfield  {author} {\bibinfo {author} {\bibfnamefont {A.~D.~K.}\
  \bibnamefont {Finck}}, \bibinfo {author} {\bibfnamefont {D.~J.}\ \bibnamefont
  {Van~Harlingen}}, \bibinfo {author} {\bibfnamefont {P.~K.}\ \bibnamefont
  {Mohseni}}, \bibinfo {author} {\bibfnamefont {K.}~\bibnamefont {Jung}}, \
  and\ \bibinfo {author} {\bibfnamefont {X.}~\bibnamefont {Li}},\ }\href
  {\doibase 10.1103/PhysRevLett.110.126406} {\bibfield  {journal} {\bibinfo
  {journal} {Phys. Rev. Lett.}\ }\textbf {\bibinfo {volume} {110}},\ \bibinfo
  {pages} {126406} (\bibinfo {year} {2013})}\BibitemShut {NoStop}%
\bibitem [{\citenamefont {Lee}\ \emph {et~al.}(2014)\citenamefont {Lee},
  \citenamefont {Jiang}, \citenamefont {Houzet}, \citenamefont {Aguado},
  \citenamefont {Lieber},\ and\ \citenamefont {De~Franceschi}}]{Lee2014:NN}%
  \BibitemOpen
  \bibfield  {author} {\bibinfo {author} {\bibfnamefont {E.~J.~H.}\
  \bibnamefont {Lee}}, \bibinfo {author} {\bibfnamefont {X.}~\bibnamefont
  {Jiang}}, \bibinfo {author} {\bibfnamefont {M.}~\bibnamefont {Houzet}},
  \bibinfo {author} {\bibfnamefont {R.}~\bibnamefont {Aguado}}, \bibinfo
  {author} {\bibfnamefont {C.~M.}\ \bibnamefont {Lieber}}, \ and\ \bibinfo
  {author} {\bibfnamefont {S.}~\bibnamefont {De~Franceschi}},\ }\href {\doibase
  10.1038/nnano.2013.267} {\bibfield  {journal} {\bibinfo  {journal} {Nat.
  Nanotechnol.}\ }\textbf {\bibinfo {volume} {9}},\ \bibinfo {pages} {79}
  (\bibinfo {year} {2014})}\BibitemShut {NoStop}%
\bibitem [{\citenamefont {Liu}\ \emph {et~al.}(2012)\citenamefont {Liu},
  \citenamefont {Potter}, \citenamefont {Law},\ and\ \citenamefont
  {Lee}}]{Liu2012:PRL}%
  \BibitemOpen
  \bibfield  {author} {\bibinfo {author} {\bibfnamefont {J.}~\bibnamefont
  {Liu}}, \bibinfo {author} {\bibfnamefont {A.~C.}\ \bibnamefont {Potter}},
  \bibinfo {author} {\bibfnamefont {K.~T.}\ \bibnamefont {Law}}, \ and\
  \bibinfo {author} {\bibfnamefont {P.~A.}\ \bibnamefont {Lee}},\ }\href
  {\doibase 10.1103/PhysRevLett.109.267002} {\bibfield  {journal} {\bibinfo
  {journal} {Phys. Rev. Lett.}\ }\textbf {\bibinfo {volume} {109}},\ \bibinfo
  {pages} {267002} (\bibinfo {year} {2012})}\BibitemShut {NoStop}%
\bibitem [{\citenamefont {Bagrets}\ and\ \citenamefont
  {Altland}(2012)}]{Bagrets2012:PRL}%
  \BibitemOpen
  \bibfield  {author} {\bibinfo {author} {\bibfnamefont {D.}~\bibnamefont
  {Bagrets}}\ and\ \bibinfo {author} {\bibfnamefont {A.}~\bibnamefont
  {Altland}},\ }\href {\doibase 10.1103/PhysRevLett.109.227005} {\bibfield
  {journal} {\bibinfo  {journal} {Phys. Rev. Lett.}\ }\textbf {\bibinfo
  {volume} {109}},\ \bibinfo {pages} {227005} (\bibinfo {year}
  {2012})}\BibitemShut {NoStop}%
\bibitem [{\citenamefont {Kells}\ \emph {et~al.}(2012)\citenamefont {Kells},
  \citenamefont {Meidan},\ and\ \citenamefont {Brouwer}}]{Kells2012:PRB}%
  \BibitemOpen
  \bibfield  {author} {\bibinfo {author} {\bibfnamefont {G.}~\bibnamefont
  {Kells}}, \bibinfo {author} {\bibfnamefont {D.}~\bibnamefont {Meidan}}, \
  and\ \bibinfo {author} {\bibfnamefont {P.~W.}\ \bibnamefont {Brouwer}},\
  }\href {\doibase 10.1103/PhysRevB.86.100503} {\bibfield  {journal} {\bibinfo
  {journal} {Phys. Rev. B}\ }\textbf {\bibinfo {volume} {86}},\ \bibinfo
  {pages} {100503} (\bibinfo {year} {2012})}\BibitemShut {NoStop}%
\bibitem [{\citenamefont {Roy}\ \emph {et~al.}(2013)\citenamefont {Roy},
  \citenamefont {Bondyopadhaya},\ and\ \citenamefont {Tewari}}]{Roy2013:PRB}%
  \BibitemOpen
  \bibfield  {author} {\bibinfo {author} {\bibfnamefont {D.}~\bibnamefont
  {Roy}}, \bibinfo {author} {\bibfnamefont {N.}~\bibnamefont {Bondyopadhaya}},
  \ and\ \bibinfo {author} {\bibfnamefont {S.}~\bibnamefont {Tewari}},\ }\href
  {\doibase 10.1103/PhysRevB.88.020502} {\bibfield  {journal} {\bibinfo
  {journal} {Phys. Rev. B}\ }\textbf {\bibinfo {volume} {88}},\ \bibinfo
  {pages} {020502} (\bibinfo {year} {2013})}\BibitemShut {NoStop}%
\bibitem [{\citenamefont {Peng}\ \emph {et~al.}(2015)\citenamefont {Peng},
  \citenamefont {Pientka}, \citenamefont {Glazman},\ and\ \citenamefont {von
  Oppen}}]{Peng2015:PRL}%
  \BibitemOpen
  \bibfield  {author} {\bibinfo {author} {\bibfnamefont {Y.}~\bibnamefont
  {Peng}}, \bibinfo {author} {\bibfnamefont {F.}~\bibnamefont {Pientka}},
  \bibinfo {author} {\bibfnamefont {L.~I.}\ \bibnamefont {Glazman}}, \ and\
  \bibinfo {author} {\bibfnamefont {F.}~\bibnamefont {von Oppen}},\ }\href
  {\doibase 10.1103/PhysRevLett.114.106801} {\bibfield  {journal} {\bibinfo
  {journal} {Phys. Rev. Lett.}\ }\textbf {\bibinfo {volume} {114}},\ \bibinfo
  {pages} {106801} (\bibinfo {year} {2015})}\BibitemShut {NoStop}%
\bibitem [{\citenamefont {Das~Sarma}\ \emph {et~al.}(2012)\citenamefont
  {Das~Sarma}, \citenamefont {Sau},\ and\ \citenamefont
  {Stanescu}}]{DasSarma2012:PRB}%
  \BibitemOpen
  \bibfield  {author} {\bibinfo {author} {\bibfnamefont {S.}~\bibnamefont
  {Das~Sarma}}, \bibinfo {author} {\bibfnamefont {J.~D.}\ \bibnamefont {Sau}},
  \ and\ \bibinfo {author} {\bibfnamefont {T.~D.}\ \bibnamefont {Stanescu}},\
  }\href {\doibase 10.1103/PhysRevB.86.220506} {\bibfield  {journal} {\bibinfo
  {journal} {Phys. Rev. B}\ }\textbf {\bibinfo {volume} {86}},\ \bibinfo
  {pages} {220506} (\bibinfo {year} {2012})}\BibitemShut {NoStop}%
\bibitem [{\citenamefont {Appelbaum}(2013)}]{Appelbaum2013:APL}%
  \BibitemOpen
  \bibfield  {author} {\bibinfo {author} {\bibfnamefont {I.}~\bibnamefont
  {Appelbaum}},\ }\href {\doibase http://dx.doi.org/10.1063/1.4821748}
  {\bibfield  {journal} {\bibinfo  {journal} {Appl. Phys. Lett.}\ }\textbf
  {\bibinfo {volume} {103}},\ \bibinfo {eid} {122604} (\bibinfo {year}
  {2013})}\BibitemShut {NoStop}%
\bibitem [{\citenamefont {Vernek}\ \emph {et~al.}(2014)\citenamefont {Vernek},
  \citenamefont {Penteado}, \citenamefont {Seridonio},\ and\ \citenamefont
  {Egues}}]{Vernek2014:PRB}%
  \BibitemOpen
  \bibfield  {author} {\bibinfo {author} {\bibfnamefont {E.}~\bibnamefont
  {Vernek}}, \bibinfo {author} {\bibfnamefont {P.~H.}\ \bibnamefont
  {Penteado}}, \bibinfo {author} {\bibfnamefont {A.~C.}\ \bibnamefont
  {Seridonio}}, \ and\ \bibinfo {author} {\bibfnamefont {J.~C.}\ \bibnamefont
  {Egues}},\ }\href {\doibase 10.1103/PhysRevB.89.165314} {\bibfield  {journal}
  {\bibinfo  {journal} {Phys. Rev. B}\ }\textbf {\bibinfo {volume} {89}},\
  \bibinfo {pages} {165314} (\bibinfo {year} {2014})}\BibitemShut {NoStop}%
\bibitem [{\citenamefont {Ben-Shach}\ \emph {et~al.}(2015)\citenamefont
  {Ben-Shach}, \citenamefont {Haim}, \citenamefont {Appelbaum}, \citenamefont
  {Oreg}, \citenamefont {Yacoby},\ and\ \citenamefont
  {Halperin}}]{BenShach2015:PRB}%
  \BibitemOpen
  \bibfield  {author} {\bibinfo {author} {\bibfnamefont {G.}~\bibnamefont
  {Ben-Shach}}, \bibinfo {author} {\bibfnamefont {A.}~\bibnamefont {Haim}},
  \bibinfo {author} {\bibfnamefont {I.}~\bibnamefont {Appelbaum}}, \bibinfo
  {author} {\bibfnamefont {Y.}~\bibnamefont {Oreg}}, \bibinfo {author}
  {\bibfnamefont {A.}~\bibnamefont {Yacoby}}, \ and\ \bibinfo {author}
  {\bibfnamefont {B.~I.}\ \bibnamefont {Halperin}},\ }\href {\doibase
  10.1103/PhysRevB.91.045403} {\bibfield  {journal} {\bibinfo  {journal} {Phys.
  Rev. B}\ }\textbf {\bibinfo {volume} {91}},\ \bibinfo {pages} {045403}
  (\bibinfo {year} {2015})}\BibitemShut {NoStop}%
\bibitem [{\citenamefont {Scharf}\ and\ \citenamefont {\ifmmode \check{Z}\else
  \v{Z}\fi{}uti\ifmmode~\acute{c}\else \'{c}\fi{}}(2015)}]{Scharf2015:PRB}%
  \BibitemOpen
  \bibfield  {author} {\bibinfo {author} {\bibfnamefont {B.}~\bibnamefont
  {Scharf}}\ and\ \bibinfo {author} {\bibfnamefont {I.}~\bibnamefont {\ifmmode
  \check{Z}\else \v{Z}\fi{}uti\ifmmode~\acute{c}\else \'{c}\fi{}}},\ }\href
  {\doibase 10.1103/PhysRevB.91.144505} {\bibfield  {journal} {\bibinfo
  {journal} {Phys. Rev. B}\ }\textbf {\bibinfo {volume} {91}},\ \bibinfo
  {pages} {144505} (\bibinfo {year} {2015})}\BibitemShut {NoStop}%
\bibitem [{\citenamefont {Kim}\ \emph {et~al.}(2015)\citenamefont {Kim},
  \citenamefont {Tewari},\ and\ \citenamefont {Tserkovnyak}}]{Kim2015:PRB}%
  \BibitemOpen
  \bibfield  {author} {\bibinfo {author} {\bibfnamefont {S.~K.}\ \bibnamefont
  {Kim}}, \bibinfo {author} {\bibfnamefont {S.}~\bibnamefont {Tewari}}, \ and\
  \bibinfo {author} {\bibfnamefont {Y.}~\bibnamefont {Tserkovnyak}},\ }\href
  {\doibase 10.1103/PhysRevB.92.020412} {\bibfield  {journal} {\bibinfo
  {journal} {Phys. Rev. B}\ }\textbf {\bibinfo {volume} {92}},\ \bibinfo
  {pages} {020412} (\bibinfo {year} {2015})}\BibitemShut {NoStop}%
\bibitem [{\citenamefont {Sau}\ \emph {et~al.}(2011)\citenamefont {Sau},
  \citenamefont {Clarke},\ and\ \citenamefont {Tewari}}]{Sau2011:PRB}%
  \BibitemOpen
  \bibfield  {author} {\bibinfo {author} {\bibfnamefont {J.~D.}\ \bibnamefont
  {Sau}}, \bibinfo {author} {\bibfnamefont {D.~J.}\ \bibnamefont {Clarke}}, \
  and\ \bibinfo {author} {\bibfnamefont {S.}~\bibnamefont {Tewari}},\ }\href
  {\doibase 10.1103/PhysRevB.84.094505} {\bibfield  {journal} {\bibinfo
  {journal} {Phys. Rev. B}\ }\textbf {\bibinfo {volume} {84}},\ \bibinfo
  {pages} {094505} (\bibinfo {year} {2011})}\BibitemShut {NoStop}%
\bibitem [{\citenamefont {Clarke}\ \emph {et~al.}(2011)\citenamefont {Clarke},
  \citenamefont {Sau},\ and\ \citenamefont {Tewari}}]{Clarke2011:PRB}%
  \BibitemOpen
  \bibfield  {author} {\bibinfo {author} {\bibfnamefont {D.~J.}\ \bibnamefont
  {Clarke}}, \bibinfo {author} {\bibfnamefont {J.~D.}\ \bibnamefont {Sau}}, \
  and\ \bibinfo {author} {\bibfnamefont {S.}~\bibnamefont {Tewari}},\ }\href
  {\doibase 10.1103/PhysRevB.84.035120} {\bibfield  {journal} {\bibinfo
  {journal} {Phys. Rev. B}\ }\textbf {\bibinfo {volume} {84}},\ \bibinfo
  {pages} {035120} (\bibinfo {year} {2011})}\BibitemShut {NoStop}%
\bibitem [{\citenamefont {Halperin}\ \emph {et~al.}(2012)\citenamefont
  {Halperin}, \citenamefont {Oreg}, \citenamefont {Stern}, \citenamefont
  {Refael}, \citenamefont {Alicea},\ and\ \citenamefont {von
  Oppen}}]{Halperin2012:PRB}%
  \BibitemOpen
  \bibfield  {author} {\bibinfo {author} {\bibfnamefont {B.~I.}\ \bibnamefont
  {Halperin}}, \bibinfo {author} {\bibfnamefont {Y.}~\bibnamefont {Oreg}},
  \bibinfo {author} {\bibfnamefont {A.}~\bibnamefont {Stern}}, \bibinfo
  {author} {\bibfnamefont {G.}~\bibnamefont {Refael}}, \bibinfo {author}
  {\bibfnamefont {J.}~\bibnamefont {Alicea}}, \ and\ \bibinfo {author}
  {\bibfnamefont {F.}~\bibnamefont {von Oppen}},\ }\href {\doibase
  10.1103/PhysRevB.85.144501} {\bibfield  {journal} {\bibinfo  {journal} {Phys.
  Rev. B}\ }\textbf {\bibinfo {volume} {85}},\ \bibinfo {pages} {144501}
  (\bibinfo {year} {2012})}\BibitemShut {NoStop}%
\bibitem [{\citenamefont {Klinovaja}\ and\ \citenamefont
  {Loss}(2013)}]{Klinovaja2013:PRX}%
  \BibitemOpen
  \bibfield  {author} {\bibinfo {author} {\bibfnamefont {J.}~\bibnamefont
  {Klinovaja}}\ and\ \bibinfo {author} {\bibfnamefont {D.}~\bibnamefont
  {Loss}},\ }\href {\doibase 10.1103/PhysRevX.3.011008} {\bibfield  {journal}
  {\bibinfo  {journal} {Phys. Rev. X}\ }\textbf {\bibinfo {volume} {3}},\
  \bibinfo {pages} {011008} (\bibinfo {year} {2013})}\BibitemShut {NoStop}%
\bibitem [{\citenamefont {Potter}\ and\ \citenamefont
  {Lee}(2010)}]{Potter2010:PRL}%
  \BibitemOpen
  \bibfield  {author} {\bibinfo {author} {\bibfnamefont {A.~C.}\ \bibnamefont
  {Potter}}\ and\ \bibinfo {author} {\bibfnamefont {P.~A.}\ \bibnamefont
  {Lee}},\ }\href {\doibase 10.1103/PhysRevLett.105.227003} {\bibfield
  {journal} {\bibinfo  {journal} {Phys. Rev. Lett.}\ }\textbf {\bibinfo
  {volume} {105}},\ \bibinfo {pages} {227003} (\bibinfo {year}
  {2010})}\BibitemShut {NoStop}%
\bibitem [{\citenamefont {Sedlmayr}\ and\ \citenamefont
  {Bena}(2015)}]{Sedlmayr2015:PRB}%
  \BibitemOpen
  \bibfield  {author} {\bibinfo {author} {\bibfnamefont {N.}~\bibnamefont
  {Sedlmayr}}\ and\ \bibinfo {author} {\bibfnamefont {C.}~\bibnamefont
  {Bena}},\ }\href {\doibase 10.1103/PhysRevB.92.115115} {\bibfield  {journal}
  {\bibinfo  {journal} {Phys. Rev. B}\ }\textbf {\bibinfo {volume} {92}},\
  \bibinfo {pages} {115115} (\bibinfo {year} {2015})}\BibitemShut {NoStop}%
\bibitem [{\citenamefont {Kjaergaard}\ \emph {et~al.}(2012)\citenamefont
  {Kjaergaard}, \citenamefont {W\"olms},\ and\ \citenamefont
  {Flensberg}}]{Kjaergaard2012:PRB}%
  \BibitemOpen
  \bibfield  {author} {\bibinfo {author} {\bibfnamefont {M.}~\bibnamefont
  {Kjaergaard}}, \bibinfo {author} {\bibfnamefont {K.}~\bibnamefont {W\"olms}},
  \ and\ \bibinfo {author} {\bibfnamefont {K.}~\bibnamefont {Flensberg}},\
  }\href {\doibase 10.1103/PhysRevB.85.020503} {\bibfield  {journal} {\bibinfo
  {journal} {Phys. Rev. B}\ }\textbf {\bibinfo {volume} {85}},\ \bibinfo
  {pages} {020503} (\bibinfo {year} {2012})}\BibitemShut {NoStop}%
\bibitem [{\citenamefont {Pientka}\ \emph {et~al.}(2013)\citenamefont
  {Pientka}, \citenamefont {Glazman},\ and\ \citenamefont {von
  Oppen}}]{Pientka2013:PRB}%
  \BibitemOpen
  \bibfield  {author} {\bibinfo {author} {\bibfnamefont {F.}~\bibnamefont
  {Pientka}}, \bibinfo {author} {\bibfnamefont {L.~I.}\ \bibnamefont
  {Glazman}}, \ and\ \bibinfo {author} {\bibfnamefont {F.}~\bibnamefont {von
  Oppen}},\ }\href {\doibase 10.1103/PhysRevB.88.155420} {\bibfield  {journal}
  {\bibinfo  {journal} {Phys. Rev. B}\ }\textbf {\bibinfo {volume} {88}},\
  \bibinfo {pages} {155420} (\bibinfo {year} {2013})}\BibitemShut {NoStop}%
\bibitem [{\citenamefont {Tsymbal}\ and\ \citenamefont
  {\v{Z}uti\'c}(2011)}]{Tsymbal2011}%
  \BibitemOpen
  \bibinfo {editor} {\bibfnamefont {E.~Y.}\ \bibnamefont {Tsymbal}}\ and\
  \bibinfo {editor} {\bibfnamefont {I.}~\bibnamefont {\v{Z}uti\'c}},\ eds.,\
  \href@noop {} {\emph {\bibinfo {title} {Handbook of Spin Transport and
  Magnetism}}}\ (\bibinfo  {publisher} {CRC Press, New York},\ \bibinfo {year}
  {2011})\BibitemShut {NoStop}%
\bibitem [{SM()}]{SM}%
  \BibitemOpen
  \href@noop {} {}\bibinfo {note} {See Supplemental Material, which
  includes~[48-51].}\BibitemShut {Stop}%
\bibitem [{\citenamefont {Ramachandran}\ and\ \citenamefont
  {Varoquaux}(2011)}]{Ramachandran2011:CSE}%
  \BibitemOpen
  \bibfield  {author} {\bibinfo {author} {\bibfnamefont {P.}~\bibnamefont
  {Ramachandran}}\ and\ \bibinfo {author} {\bibfnamefont {G.}~\bibnamefont
  {Varoquaux}},\ }\href@noop {} {\bibfield  {journal} {\bibinfo  {journal}
  {Comput. Sci. Eng.}\ }\textbf {\bibinfo {volume} {13}},\ \bibinfo {pages}
  {40} (\bibinfo {year} {2011})}\BibitemShut {NoStop}%
\bibitem [{\citenamefont {Chartrand}\ and\ \citenamefont
  {Zhang}(2004)}]{Chartrand2004}%
  \BibitemOpen
  \bibfield  {author} {\bibinfo {author} {\bibfnamefont {G.}~\bibnamefont
  {Chartrand}}\ and\ \bibinfo {author} {\bibfnamefont {P.}~\bibnamefont
  {Zhang}},\ }\href@noop {} {\emph {\bibinfo {title} {Introduction to Graph
  Theory}}}\ (\bibinfo  {publisher} {McGraw-Hill, New York},\ \bibinfo {year}
  {2004})\BibitemShut {NoStop}%
\bibitem [{\citenamefont {Zudov}\ \emph {et~al.}(2002)\citenamefont {Zudov},
  \citenamefont {Kono}, \citenamefont {Matsuda}, \citenamefont {Ikaida},
  \citenamefont {Miura}, \citenamefont {Munekata}, \citenamefont {Sanders},
  \citenamefont {Sun},\ and\ \citenamefont {Stanton}}]{Zudov2002:PRB}%
  \BibitemOpen
  \bibfield  {author} {\bibinfo {author} {\bibfnamefont {M.~A.}\ \bibnamefont
  {Zudov}}, \bibinfo {author} {\bibfnamefont {J.}~\bibnamefont {Kono}},
  \bibinfo {author} {\bibfnamefont {Y.~H.}\ \bibnamefont {Matsuda}}, \bibinfo
  {author} {\bibfnamefont {T.}~\bibnamefont {Ikaida}}, \bibinfo {author}
  {\bibfnamefont {N.}~\bibnamefont {Miura}}, \bibinfo {author} {\bibfnamefont
  {H.}~\bibnamefont {Munekata}}, \bibinfo {author} {\bibfnamefont {G.~D.}\
  \bibnamefont {Sanders}}, \bibinfo {author} {\bibfnamefont {Y.}~\bibnamefont
  {Sun}}, \ and\ \bibinfo {author} {\bibfnamefont {C.~J.}\ \bibnamefont
  {Stanton}},\ }\href {\doibase 10.1103/PhysRevB.66.161307} {\bibfield
  {journal} {\bibinfo  {journal} {Phys. Rev. B}\ }\textbf {\bibinfo {volume}
  {66}},\ \bibinfo {pages} {161307} (\bibinfo {year} {2002})}\BibitemShut
  {NoStop}%
\bibitem [{\citenamefont {Fabian}\ \emph {et~al.}(2007)\citenamefont {Fabian},
  \citenamefont {Matos-Abiague}, \citenamefont {Ertler}, \citenamefont
  {Stano},\ and\ \citenamefont {\ifmmode \check{Z}\else
  \v{Z}\fi{}uti\ifmmode~\acute{c}\else \'{c}\fi{}}}]{Fabian2007:APS}%
  \BibitemOpen
  \bibfield  {author} {\bibinfo {author} {\bibfnamefont {J.}~\bibnamefont
  {Fabian}}, \bibinfo {author} {\bibfnamefont {A.}~\bibnamefont
  {Matos-Abiague}}, \bibinfo {author} {\bibfnamefont {C.}~\bibnamefont
  {Ertler}}, \bibinfo {author} {\bibfnamefont {P.}~\bibnamefont {Stano}}, \
  and\ \bibinfo {author} {\bibfnamefont {I.}~\bibnamefont {\ifmmode
  \check{Z}\else \v{Z}\fi{}uti\ifmmode~\acute{c}\else \'{c}\fi{}}},\
  }\href@noop {} {\bibfield  {journal} {\bibinfo  {journal} {Acta Phys. Slov.}\
  }\textbf {\bibinfo {volume} {57}},\ \bibinfo {pages} {565} (\bibinfo {year}
  {2007})}\BibitemShut {NoStop}%
\bibitem [{\citenamefont {Braunecker}\ \emph {et~al.}(2010)\citenamefont
  {Braunecker}, \citenamefont {Japaridze}, \citenamefont {Klinovaja},\ and\
  \citenamefont {Loss}}]{Braunecker2010:PRB}%
  \BibitemOpen
  \bibfield  {author} {\bibinfo {author} {\bibfnamefont {B.}~\bibnamefont
  {Braunecker}}, \bibinfo {author} {\bibfnamefont {G.~I.}\ \bibnamefont
  {Japaridze}}, \bibinfo {author} {\bibfnamefont {J.}~\bibnamefont
  {Klinovaja}}, \ and\ \bibinfo {author} {\bibfnamefont {D.}~\bibnamefont
  {Loss}},\ }\href {\doibase 10.1103/PhysRevB.82.045127} {\bibfield  {journal}
  {\bibinfo  {journal} {Phys. Rev. B}\ }\textbf {\bibinfo {volume} {82}},\
  \bibinfo {pages} {045127} (\bibinfo {year} {2010})}\BibitemShut {NoStop}%
\bibitem [{\citenamefont {Alicea}(2010)}]{Alicea2010:PRB}%
  \BibitemOpen
  \bibfield  {author} {\bibinfo {author} {\bibfnamefont {J.}~\bibnamefont
  {Alicea}},\ }\href {\doibase 10.1103/PhysRevB.81.125318} {\bibfield
  {journal} {\bibinfo  {journal} {Phys. Rev. B}\ }\textbf {\bibinfo {volume}
  {81}},\ \bibinfo {pages} {125318} (\bibinfo {year} {2010})}\BibitemShut
  {NoStop}%
\bibitem [{\citenamefont {Sau}\ \emph {et~al.}(2010{\natexlab{b}})\citenamefont
  {Sau}, \citenamefont {Tewari}, \citenamefont {Lutchyn}, \citenamefont
  {Stanescu},\ and\ \citenamefont {Das~Sarma}}]{Sau2010:PRB}%
  \BibitemOpen
  \bibfield  {author} {\bibinfo {author} {\bibfnamefont {J.~D.}\ \bibnamefont
  {Sau}}, \bibinfo {author} {\bibfnamefont {S.}~\bibnamefont {Tewari}},
  \bibinfo {author} {\bibfnamefont {R.~M.}\ \bibnamefont {Lutchyn}}, \bibinfo
  {author} {\bibfnamefont {T.~D.}\ \bibnamefont {Stanescu}}, \ and\ \bibinfo
  {author} {\bibfnamefont {S.}~\bibnamefont {Das~Sarma}},\ }\href {\doibase
  10.1103/PhysRevB.82.214509} {\bibfield  {journal} {\bibinfo  {journal} {Phys.
  Rev. B}\ }\textbf {\bibinfo {volume} {82}},\ \bibinfo {pages} {214509}
  (\bibinfo {year} {2010}{\natexlab{b}})}\BibitemShut {NoStop}%
\bibitem [{\citenamefont {Horn}\ and\ \citenamefont
  {Johnson}(1991)}]{Horn1991}%
  \BibitemOpen
  \bibfield  {author} {\bibinfo {author} {\bibfnamefont {R.~A.}\ \bibnamefont
  {Horn}}\ and\ \bibinfo {author} {\bibfnamefont {C.~R.}\ \bibnamefont
  {Johnson}},\ }\href {http://dx.doi.org/10.1017/CBO9780511840371} {\emph
  {\bibinfo {title} {Topics in Matrix Analysis}}}\ (\bibinfo  {publisher}
  {Cambridge University Press},\ \bibinfo {year} {1991})\BibitemShut {NoStop}%
\bibitem [{Note2()}]{Note2}%
  \BibitemOpen
  \bibinfo {note} {Superconducting proximity effects and MBSs in (Cd,Mn) Te are
  being explored by L. Rokhinson (private comm.).}\BibitemShut {Stop}%
\bibitem [{\citenamefont {Betthausen}\ \emph {et~al.}(2012)\citenamefont
  {Betthausen}, \citenamefont {Dollinger}, \citenamefont {Saarikoski},
  \citenamefont {Kolkovsky}, \citenamefont {Karczewski}, \citenamefont
  {Wojtowicz}, \citenamefont {Richter},\ and\ \citenamefont
  {Weiss}}]{Betthausen2012:Science}%
  \BibitemOpen
  \bibfield  {author} {\bibinfo {author} {\bibfnamefont {C.}~\bibnamefont
  {Betthausen}}, \bibinfo {author} {\bibfnamefont {T.}~\bibnamefont
  {Dollinger}}, \bibinfo {author} {\bibfnamefont {H.}~\bibnamefont
  {Saarikoski}}, \bibinfo {author} {\bibfnamefont {V.}~\bibnamefont
  {Kolkovsky}}, \bibinfo {author} {\bibfnamefont {G.}~\bibnamefont
  {Karczewski}}, \bibinfo {author} {\bibfnamefont {T.}~\bibnamefont
  {Wojtowicz}}, \bibinfo {author} {\bibfnamefont {K.}~\bibnamefont {Richter}},
  \ and\ \bibinfo {author} {\bibfnamefont {D.}~\bibnamefont {Weiss}},\ }\href
  {\doibase 10.1126/science.1221350} {\bibfield  {journal} {\bibinfo  {journal}
  {Science}\ }\textbf {\bibinfo {volume} {337}},\ \bibinfo {pages} {324}
  (\bibinfo {year} {2012})}\BibitemShut {NoStop}%
\bibitem [{\citenamefont {Lim}\ \emph {et~al.}(2012)\citenamefont {Lim},
  \citenamefont {Serra}, \citenamefont {L\'opez},\ and\ \citenamefont
  {Aguado}}]{Lim2012:PRB}%
  \BibitemOpen
  \bibfield  {author} {\bibinfo {author} {\bibfnamefont {J.~S.}\ \bibnamefont
  {Lim}}, \bibinfo {author} {\bibfnamefont {L.~m.~c.}\ \bibnamefont {Serra}},
  \bibinfo {author} {\bibfnamefont {R.}~\bibnamefont {L\'opez}}, \ and\
  \bibinfo {author} {\bibfnamefont {R.}~\bibnamefont {Aguado}},\ }\href
  {\doibase 10.1103/PhysRevB.86.121103} {\bibfield  {journal} {\bibinfo
  {journal} {Phys. Rev. B}\ }\textbf {\bibinfo {volume} {86}},\ \bibinfo
  {pages} {121103} (\bibinfo {year} {2012})}\BibitemShut {NoStop}%
\bibitem [{\citenamefont {Prada}\ \emph {et~al.}(2012)\citenamefont {Prada},
  \citenamefont {San-Jose},\ and\ \citenamefont {Aguado}}]{Prada2012:PRB}%
  \BibitemOpen
  \bibfield  {author} {\bibinfo {author} {\bibfnamefont {E.}~\bibnamefont
  {Prada}}, \bibinfo {author} {\bibfnamefont {P.}~\bibnamefont {San-Jose}}, \
  and\ \bibinfo {author} {\bibfnamefont {R.}~\bibnamefont {Aguado}},\ }\href
  {\doibase 10.1103/PhysRevB.86.180503} {\bibfield  {journal} {\bibinfo
  {journal} {Phys. Rev. B}\ }\textbf {\bibinfo {volume} {86}},\ \bibinfo
  {pages} {180503} (\bibinfo {year} {2012})}\BibitemShut {NoStop}%
\bibitem [{\citenamefont {Rainis}\ \emph {et~al.}(2013)\citenamefont {Rainis},
  \citenamefont {Trifunovic}, \citenamefont {Klinovaja},\ and\ \citenamefont
  {Loss}}]{Rainis2013:PRB}%
  \BibitemOpen
  \bibfield  {author} {\bibinfo {author} {\bibfnamefont {D.}~\bibnamefont
  {Rainis}}, \bibinfo {author} {\bibfnamefont {L.}~\bibnamefont {Trifunovic}},
  \bibinfo {author} {\bibfnamefont {J.}~\bibnamefont {Klinovaja}}, \ and\
  \bibinfo {author} {\bibfnamefont {D.}~\bibnamefont {Loss}},\ }\href {\doibase
  10.1103/PhysRevB.87.024515} {\bibfield  {journal} {\bibinfo  {journal} {Phys.
  Rev. B}\ }\textbf {\bibinfo {volume} {87}},\ \bibinfo {pages} {024515}
  (\bibinfo {year} {2013})}\BibitemShut {NoStop}%
\bibitem [{\citenamefont {{Hart}}\ \emph {et~al.}(2015)\citenamefont {{Hart}},
  \citenamefont {{Ren}}, \citenamefont {{Kosowsky}}, \citenamefont
  {{Ben-Shach}}, \citenamefont {{Leubner}}, \citenamefont {{Br{\"u}ne}},
  \citenamefont {{Buhmann}}, \citenamefont {{Molenkamp}}, \citenamefont
  {{Halperin}},\ and\ \citenamefont {{Yacoby}}}]{Hart2015:arXiv}%
  \BibitemOpen
  \bibfield  {author} {\bibinfo {author} {\bibfnamefont {S.}~\bibnamefont
  {{Hart}}}, \bibinfo {author} {\bibfnamefont {H.}~\bibnamefont {{Ren}}},
  \bibinfo {author} {\bibfnamefont {M.}~\bibnamefont {{Kosowsky}}}, \bibinfo
  {author} {\bibfnamefont {G.}~\bibnamefont {{Ben-Shach}}}, \bibinfo {author}
  {\bibfnamefont {P.}~\bibnamefont {{Leubner}}}, \bibinfo {author}
  {\bibfnamefont {C.}~\bibnamefont {{Br{\"u}ne}}}, \bibinfo {author}
  {\bibfnamefont {H.}~\bibnamefont {{Buhmann}}}, \bibinfo {author}
  {\bibfnamefont {L.~W.}\ \bibnamefont {{Molenkamp}}}, \bibinfo {author}
  {\bibfnamefont {B.~I.}\ \bibnamefont {{Halperin}}}, \ and\ \bibinfo {author}
  {\bibfnamefont {A.}~\bibnamefont {{Yacoby}}},\ }\href@noop {} {\bibfield
  {journal} {\bibinfo  {journal} {arXiv:1509.02940}\ } (\bibinfo {year}
  {2015})}\BibitemShut {NoStop}%
\bibitem [{\citenamefont {Xu}\ \emph {et~al.}(2014)\citenamefont {Xu},
  \citenamefont {Alidoust}, \citenamefont {Belopolski}, \citenamefont
  {Richardella}, \citenamefont {Liu}, \citenamefont {Neupane}, \citenamefont
  {Bian}, \citenamefont {Huang}, \citenamefont {Sankar}, \citenamefont {Fang},
  \citenamefont {Dellabetta}, \citenamefont {Dai}, \citenamefont {Li},
  \citenamefont {Gilbert}, \citenamefont {Chou}, \citenamefont {Samarth},\ and\
  \citenamefont {Hasan}}]{Xu2014:NP}%
  \BibitemOpen
  \bibfield  {author} {\bibinfo {author} {\bibfnamefont {S.-Y.}\ \bibnamefont
  {Xu}}, \bibinfo {author} {\bibfnamefont {N.}~\bibnamefont {Alidoust}},
  \bibinfo {author} {\bibfnamefont {I.}~\bibnamefont {Belopolski}}, \bibinfo
  {author} {\bibfnamefont {A.}~\bibnamefont {Richardella}}, \bibinfo {author}
  {\bibfnamefont {C.}~\bibnamefont {Liu}}, \bibinfo {author} {\bibfnamefont
  {M.}~\bibnamefont {Neupane}}, \bibinfo {author} {\bibfnamefont
  {G.}~\bibnamefont {Bian}}, \bibinfo {author} {\bibfnamefont {S.-H.}\
  \bibnamefont {Huang}}, \bibinfo {author} {\bibfnamefont {R.}~\bibnamefont
  {Sankar}}, \bibinfo {author} {\bibfnamefont {C.}~\bibnamefont {Fang}},
  \bibinfo {author} {\bibfnamefont {B.}~\bibnamefont {Dellabetta}}, \bibinfo
  {author} {\bibfnamefont {W.}~\bibnamefont {Dai}}, \bibinfo {author}
  {\bibfnamefont {Q.}~\bibnamefont {Li}}, \bibinfo {author} {\bibfnamefont
  {M.~J.}\ \bibnamefont {Gilbert}}, \bibinfo {author} {\bibfnamefont
  {F.}~\bibnamefont {Chou}}, \bibinfo {author} {\bibfnamefont {N.}~\bibnamefont
  {Samarth}}, \ and\ \bibinfo {author} {\bibfnamefont {M.~Z.}\ \bibnamefont
  {Hasan}},\ }\href {http://dx.doi.org/10.1038/nphys3139} {\bibfield  {journal}
  {\bibinfo  {journal} {Nat. Phys.}\ }\textbf {\bibinfo {volume} {10}},\
  \bibinfo {pages} {943} (\bibinfo {year} {2014})}\BibitemShut {NoStop}%
\bibitem [{\citenamefont {Shabani}\ \emph {et~al.}(2016)\citenamefont
  {Shabani}, \citenamefont {Kjaergaard}, \citenamefont {Suominen},
  \citenamefont {Kim}, \citenamefont {Nichele}, \citenamefont {Pakrouski},
  \citenamefont {Stankevic}, \citenamefont {Lutchyn}, \citenamefont
  {Krogstrup}, \citenamefont {Feidenhans'l}, \citenamefont {Kraemer},
  \citenamefont {Nayak}, \citenamefont {Troyer}, \citenamefont {Marcus},\ and\
  \citenamefont {Palmstr\o{}m}}]{Shabani2015:PRB}%
  \BibitemOpen
  \bibfield  {author} {\bibinfo {author} {\bibfnamefont {J.}~\bibnamefont
  {Shabani}}, \bibinfo {author} {\bibfnamefont {M.}~\bibnamefont {Kjaergaard}},
  \bibinfo {author} {\bibfnamefont {H.~J.}\ \bibnamefont {Suominen}}, \bibinfo
  {author} {\bibfnamefont {Y.}~\bibnamefont {Kim}}, \bibinfo {author}
  {\bibfnamefont {F.}~\bibnamefont {Nichele}}, \bibinfo {author} {\bibfnamefont
  {K.}~\bibnamefont {Pakrouski}}, \bibinfo {author} {\bibfnamefont
  {T.}~\bibnamefont {Stankevic}}, \bibinfo {author} {\bibfnamefont {R.~M.}\
  \bibnamefont {Lutchyn}}, \bibinfo {author} {\bibfnamefont {P.}~\bibnamefont
  {Krogstrup}}, \bibinfo {author} {\bibfnamefont {R.}~\bibnamefont
  {Feidenhans'l}}, \bibinfo {author} {\bibfnamefont {S.}~\bibnamefont
  {Kraemer}}, \bibinfo {author} {\bibfnamefont {C.}~\bibnamefont {Nayak}},
  \bibinfo {author} {\bibfnamefont {M.}~\bibnamefont {Troyer}}, \bibinfo
  {author} {\bibfnamefont {C.~M.}\ \bibnamefont {Marcus}}, \ and\ \bibinfo
  {author} {\bibfnamefont {C.~J.}\ \bibnamefont {Palmstr\o{}m}},\ }\href
  {\doibase 10.1103/PhysRevB.93.155402} {\bibfield  {journal} {\bibinfo
  {journal} {Phys. Rev. B}\ }\textbf {\bibinfo {volume} {93}},\ \bibinfo
  {pages} {155402} (\bibinfo {year} {2016})}\BibitemShut {NoStop}%
\bibitem [{\citenamefont {{Winkler}}\ \emph {et~al.}()\citenamefont
  {{Winkler}}, \citenamefont {{Wu}}, \citenamefont {{Troyer}}, \citenamefont
  {{Krogstrup}},\ and\ \citenamefont {{Soluyanov}}}]{Winkler2016:arxiv}%
  \BibitemOpen
  \bibfield  {author} {\bibinfo {author} {\bibfnamefont {G.~W.}\ \bibnamefont
  {{Winkler}}}, \bibinfo {author} {\bibfnamefont {Q.-S.}\ \bibnamefont {{Wu}}},
  \bibinfo {author} {\bibfnamefont {M.}~\bibnamefont {{Troyer}}}, \bibinfo
  {author} {\bibfnamefont {P.}~\bibnamefont {{Krogstrup}}}, \ and\ \bibinfo
  {author} {\bibfnamefont {A.~A.}\ \bibnamefont {{Soluyanov}}},\ }\href@noop {}
  {\bibinfo  {journal} {arXiv:1602.07001}\ }\BibitemShut {NoStop}%
\bibitem [{\citenamefont {\ifmmode \check{Z}\else
  \v{Z}\fi{}uti\ifmmode~\acute{c}\else \'{c}\fi{}}\ \emph
  {et~al.}(2004)\citenamefont {\ifmmode \check{Z}\else
  \v{Z}\fi{}uti\ifmmode~\acute{c}\else \'{c}\fi{}}, \citenamefont {Fabian},\
  and\ \citenamefont {Das~Sarma}}]{Zutic2004:RMP}%
  \BibitemOpen
\bibfield  {journal} {  }\bibfield  {author} {\bibinfo {author} {\bibfnamefont
  {I.}~\bibnamefont {\ifmmode \check{Z}\else
  \v{Z}\fi{}uti\ifmmode~\acute{c}\else \'{c}\fi{}}}, \bibinfo {author}
  {\bibfnamefont {J.}~\bibnamefont {Fabian}}, \ and\ \bibinfo {author}
  {\bibfnamefont {S.}~\bibnamefont {Das~Sarma}},\ }\href {\doibase
  10.1103/RevModPhys.76.323} {\bibfield  {journal} {\bibinfo  {journal} {Rev.
  Mod. Phys.}\ }\textbf {\bibinfo {volume} {76}},\ \bibinfo {pages} {323}
  (\bibinfo {year} {2004})}\BibitemShut {NoStop}%
\bibitem [{\citenamefont {Lin}\ \emph {et~al.}(2012)\citenamefont {Lin},
  \citenamefont {Sau},\ and\ \citenamefont {Das~Sarma}}]{Lin2012:PRB}%
  \BibitemOpen
  \bibfield  {author} {\bibinfo {author} {\bibfnamefont {C.-H.}\ \bibnamefont
  {Lin}}, \bibinfo {author} {\bibfnamefont {J.~D.}\ \bibnamefont {Sau}}, \ and\
  \bibinfo {author} {\bibfnamefont {S.}~\bibnamefont {Das~Sarma}},\ }\href
  {\doibase 10.1103/PhysRevB.86.224511} {\bibfield  {journal} {\bibinfo
  {journal} {Phys. Rev. B}\ }\textbf {\bibinfo {volume} {86}},\ \bibinfo
  {pages} {224511} (\bibinfo {year} {2012})}\BibitemShut {NoStop}%
\bibitem [{\citenamefont {Fridman}\ \emph {et~al.}(2011)\citenamefont
  {Fridman}, \citenamefont {Kloc}, \citenamefont {Petrovic},\ and\
  \citenamefont {Wei}}]{Fridman2011:APL}%
  \BibitemOpen
  \bibfield  {author} {\bibinfo {author} {\bibfnamefont {I.}~\bibnamefont
  {Fridman}}, \bibinfo {author} {\bibfnamefont {C.}~\bibnamefont {Kloc}},
  \bibinfo {author} {\bibfnamefont {C.}~\bibnamefont {Petrovic}}, \ and\
  \bibinfo {author} {\bibfnamefont {J.~Y.~T.}\ \bibnamefont {Wei}},\ }\href
  {\doibase http://dx.doi.org/10.1063/1.3659412} {\bibfield  {journal}
  {\bibinfo  {journal} {Appl. Phys. Lett.}\ }\textbf {\bibinfo {volume} {99}},\
  \bibinfo {pages} {192505} (\bibinfo {year} {2011})}\BibitemShut {NoStop}%
\bibitem [{\citenamefont {Ilani}\ \emph {et~al.}(2004)\citenamefont {Ilani},
  \citenamefont {Martin}, \citenamefont {Teitelbaum}, \citenamefont {Smet},
  \citenamefont {Mahalu}, \citenamefont {Umansky},\ and\ \citenamefont
  {Yacoby}}]{Ilani2004:Nature}%
  \BibitemOpen
  \bibfield  {author} {\bibinfo {author} {\bibfnamefont {S.}~\bibnamefont
  {Ilani}}, \bibinfo {author} {\bibfnamefont {J.}~\bibnamefont {Martin}},
  \bibinfo {author} {\bibfnamefont {E.}~\bibnamefont {Teitelbaum}}, \bibinfo
  {author} {\bibfnamefont {J.~H.}\ \bibnamefont {Smet}}, \bibinfo {author}
  {\bibfnamefont {D.}~\bibnamefont {Mahalu}}, \bibinfo {author} {\bibfnamefont
  {V.}~\bibnamefont {Umansky}}, \ and\ \bibinfo {author} {\bibfnamefont
  {A.}~\bibnamefont {Yacoby}},\ }\href@noop {} {\bibfield  {journal} {\bibinfo
  {journal} {Nature}\ }\textbf {\bibinfo {volume} {427}},\ \bibinfo {pages}
  {328} (\bibinfo {year} {2004})}\BibitemShut {NoStop}%
\bibitem [{\citenamefont {Martin}\ \emph {et~al.}(2004)\citenamefont {Martin},
  \citenamefont {Ilani}, \citenamefont {Verdene}, \citenamefont {Smet},
  \citenamefont {Umansky}, \citenamefont {Mahalu}, \citenamefont {Schuh},
  \citenamefont {Abstreiter},\ and\ \citenamefont
  {Yacoby}}]{Martin2004:Science}%
  \BibitemOpen
  \bibfield  {author} {\bibinfo {author} {\bibfnamefont {J.}~\bibnamefont
  {Martin}}, \bibinfo {author} {\bibfnamefont {S.}~\bibnamefont {Ilani}},
  \bibinfo {author} {\bibfnamefont {B.}~\bibnamefont {Verdene}}, \bibinfo
  {author} {\bibfnamefont {J.}~\bibnamefont {Smet}}, \bibinfo {author}
  {\bibfnamefont {V.}~\bibnamefont {Umansky}}, \bibinfo {author} {\bibfnamefont
  {D.}~\bibnamefont {Mahalu}}, \bibinfo {author} {\bibfnamefont
  {D.}~\bibnamefont {Schuh}}, \bibinfo {author} {\bibfnamefont
  {G.}~\bibnamefont {Abstreiter}}, \ and\ \bibinfo {author} {\bibfnamefont
  {A.}~\bibnamefont {Yacoby}},\ }\href {\doibase 10.1126/science.1099950}
  {\bibfield  {journal} {\bibinfo  {journal} {Science}\ }\textbf {\bibinfo
  {volume} {305}},\ \bibinfo {pages} {980} (\bibinfo {year}
  {2004})}\BibitemShut {NoStop}%
\bibitem [{\citenamefont {Asano}\ \emph {et~al.}(2010)\citenamefont {Asano},
  \citenamefont {Tanaka},\ and\ \citenamefont {Nagaosa}}]{Asano2010:PRL}%
  \BibitemOpen
  \bibfield  {author} {\bibinfo {author} {\bibfnamefont {Y.}~\bibnamefont
  {Asano}}, \bibinfo {author} {\bibfnamefont {Y.}~\bibnamefont {Tanaka}}, \
  and\ \bibinfo {author} {\bibfnamefont {N.}~\bibnamefont {Nagaosa}},\ }\href
  {\doibase 10.1103/PhysRevLett.105.056402} {\bibfield  {journal} {\bibinfo
  {journal} {Phys. Rev. Lett.}\ }\textbf {\bibinfo {volume} {105}},\ \bibinfo
  {pages} {056402} (\bibinfo {year} {2010})}\BibitemShut {NoStop}%
\bibitem [{\citenamefont {Gangadharaiah}\ \emph {et~al.}(2011)\citenamefont
  {Gangadharaiah}, \citenamefont {Braunecker}, \citenamefont {Simon},\ and\
  \citenamefont {Loss}}]{Gangadharaiah2011:PRL}%
  \BibitemOpen
  \bibfield  {author} {\bibinfo {author} {\bibfnamefont {S.}~\bibnamefont
  {Gangadharaiah}}, \bibinfo {author} {\bibfnamefont {B.}~\bibnamefont
  {Braunecker}}, \bibinfo {author} {\bibfnamefont {P.}~\bibnamefont {Simon}}, \
  and\ \bibinfo {author} {\bibfnamefont {D.}~\bibnamefont {Loss}},\ }\href
  {\doibase 10.1103/PhysRevLett.107.036801} {\bibfield  {journal} {\bibinfo
  {journal} {Phys. Rev. Lett.}\ }\textbf {\bibinfo {volume} {107}},\ \bibinfo
  {pages} {036801} (\bibinfo {year} {2011})}\BibitemShut {NoStop}%
\bibitem [{\citenamefont {Stoudenmire}\ \emph {et~al.}(2011)\citenamefont
  {Stoudenmire}, \citenamefont {Alicea}, \citenamefont {Starykh},\ and\
  \citenamefont {Fisher}}]{Stoudenmire2011:PRB}%
  \BibitemOpen
  \bibfield  {author} {\bibinfo {author} {\bibfnamefont {E.~M.}\ \bibnamefont
  {Stoudenmire}}, \bibinfo {author} {\bibfnamefont {J.}~\bibnamefont {Alicea}},
  \bibinfo {author} {\bibfnamefont {O.~A.}\ \bibnamefont {Starykh}}, \ and\
  \bibinfo {author} {\bibfnamefont {M.~P.}\ \bibnamefont {Fisher}},\ }\href
  {\doibase 10.1103/PhysRevB.84.014503} {\bibfield  {journal} {\bibinfo
  {journal} {Phys. Rev. B}\ }\textbf {\bibinfo {volume} {84}},\ \bibinfo
  {pages} {014503} (\bibinfo {year} {2011})}\BibitemShut {NoStop}%
\bibitem [{\citenamefont {Nakosai}\ \emph {et~al.}(2013)\citenamefont
  {Nakosai}, \citenamefont {Budich}, \citenamefont {Tanaka}, \citenamefont
  {Trauzettel},\ and\ \citenamefont {Nagaosa}}]{Nakosai2013:PRL}%
  \BibitemOpen
  \bibfield  {author} {\bibinfo {author} {\bibfnamefont {S.}~\bibnamefont
  {Nakosai}}, \bibinfo {author} {\bibfnamefont {J.~C.}\ \bibnamefont {Budich}},
  \bibinfo {author} {\bibfnamefont {Y.}~\bibnamefont {Tanaka}}, \bibinfo
  {author} {\bibfnamefont {B.}~\bibnamefont {Trauzettel}}, \ and\ \bibinfo
  {author} {\bibfnamefont {N.}~\bibnamefont {Nagaosa}},\ }\href {\doibase
  10.1103/PhysRevLett.110.117002} {\bibfield  {journal} {\bibinfo  {journal}
  {Phys. Rev. Lett.}\ }\textbf {\bibinfo {volume} {110}},\ \bibinfo {pages}
  {117002} (\bibinfo {year} {2013})}\BibitemShut {NoStop}%
\bibitem [{\citenamefont {Adagideli}\ \emph {et~al.}(2014)\citenamefont
  {Adagideli}, \citenamefont {Wimmer},\ and\ \citenamefont
  {Teker}}]{Adagideli2014:PRB}%
  \BibitemOpen
  \bibfield  {author} {\bibinfo {author} {\bibfnamefont {I.}~\bibnamefont
  {Adagideli}}, \bibinfo {author} {\bibfnamefont {M.}~\bibnamefont {Wimmer}}, \
  and\ \bibinfo {author} {\bibfnamefont {A.}~\bibnamefont {Teker}},\ }\href
  {\doibase 10.1103/PhysRevB.89.144506} {\bibfield  {journal} {\bibinfo
  {journal} {Phys. Rev. B}\ }\textbf {\bibinfo {volume} {89}},\ \bibinfo
  {pages} {144506} (\bibinfo {year} {2014})}\BibitemShut {NoStop}%
\bibitem [{\citenamefont {Valentini}\ \emph {et~al.}(2014)\citenamefont
  {Valentini}, \citenamefont {Fazio},\ and\ \citenamefont
  {Taddei}}]{Valentini2014:PRB}%
  \BibitemOpen
  \bibfield  {author} {\bibinfo {author} {\bibfnamefont {S.}~\bibnamefont
  {Valentini}}, \bibinfo {author} {\bibfnamefont {R.}~\bibnamefont {Fazio}}, \
  and\ \bibinfo {author} {\bibfnamefont {F.}~\bibnamefont {Taddei}},\ }\href
  {\doibase 10.1103/PhysRevB.89.014509} {\bibfield  {journal} {\bibinfo
  {journal} {Phys. Rev. B}\ }\textbf {\bibinfo {volume} {89}},\ \bibinfo
  {pages} {014509} (\bibinfo {year} {2014})}\BibitemShut {NoStop}%
\bibitem [{\citenamefont {Kashuba}\ and\ \citenamefont
  {Timm}(2015)}]{Kashuba2015:PRL}%
  \BibitemOpen
  \bibfield  {author} {\bibinfo {author} {\bibfnamefont {O.}~\bibnamefont
  {Kashuba}}\ and\ \bibinfo {author} {\bibfnamefont {C.}~\bibnamefont {Timm}},\
  }\href {\doibase 10.1103/PhysRevLett.114.116801} {\bibfield  {journal}
  {\bibinfo  {journal} {Phys. Rev. Lett.}\ }\textbf {\bibinfo {volume} {114}},\
  \bibinfo {pages} {116801} (\bibinfo {year} {2015})}\BibitemShut {NoStop}%
\bibitem [{\citenamefont {Kovalev}\ \emph {et~al.}(2014)\citenamefont
  {Kovalev}, \citenamefont {De},\ and\ \citenamefont
  {Shtengel}}]{Kovalev2014:PRL}%
  \BibitemOpen
  \bibfield  {author} {\bibinfo {author} {\bibfnamefont {A.~A.}\ \bibnamefont
  {Kovalev}}, \bibinfo {author} {\bibfnamefont {A.}~\bibnamefont {De}}, \ and\
  \bibinfo {author} {\bibfnamefont {K.}~\bibnamefont {Shtengel}},\ }\href
  {\doibase 10.1103/PhysRevLett.112.106402} {\bibfield  {journal} {\bibinfo
  {journal} {Phys. Rev. Lett.}\ }\textbf {\bibinfo {volume} {112}},\ \bibinfo
  {pages} {106402} (\bibinfo {year} {2014})}\BibitemShut {NoStop}%
\end{thebibliography}%

\end{document}